%% file: sample631.tex
\documentclass{aastex631}

\newcommand*{\ditto}{\texttt{"}}

\received{2024 Jul 9}
\revised{2024 Sep 30}
\accepted{2024 Oct 22}

\submitjournal{PSJ}

\begin{document}

\title{Profiling Near-Surface Winds on Mars Using Attitude Data from Mars 2020 Ingenuity}

\author[0000-0002-9495-9700]{Brian Jackson}
\affiliation{Department of Physics, Boise State University, 1910 University Drive, Boise ID 83725-1570 USA}
\affiliation{Carl Sagan Center, SETI Institute, Mountain View, CA, United States}

\author[0000-0001-8116-4901]{Lori Fenton}
\affiliation{Carl Sagan Center, SETI Institute, Mountain View, CA, United States}

\author[0000-0002-4846-8468]{Travis Brown}
\affiliation{Jet Propulsion Laboratory, California Institute of Technology, Pasadena CA 91109 USA}

\author[0000-0002-1677-6327]{Asier Munguira}
\affiliation{Universidad del País Vasco UPV/EHU, Bilbao, Spain}

\author{German Martinez}
\affiliation{Lunar and Planetary Institute, 3600 Bay Area Boulevard, Office 2029, Houston, TX 77058}
\affiliation{Centro de Astrobiologia (CAB, CSIC-INTA) and National Institute for Aerospace Technology (INTA), Madrid, Spain}

\author{Claire Newman}
\affiliation{Aeolis Research, Tucson, AZ, USA}

\author{Daniel Vi\'{u}dez-Moreiras}
\affiliation{Centro de Astrobiologia (CAB, CSIC-INTA) and National Institute for Aerospace Technology (INTA), Madrid, Spain}

\author[0000-0002-1928-2293]{Matthew Golombek}
\affiliation{Jet Propulsion Laboratory, California Institute of Technology, Pasadena CA 91109 USA}

\author[0000-0001-8528-4644]{Ralph Lorenz}
\affiliation{Johns Hopkins Applied Physics Laboratory, 1100 Johns Hopkins Road, Laurel, MD, USA }

\author{Mark D.~Paton}
\affiliation{Finnish Meteorological Institute, PO Box 503, FIN-00101 Helsinki, Finland}

\author{Dylan Conway}
\affiliation{Jet Propulsion Laboratory, California Institute of Technology, Pasadena CA 91109 USA}

\begin{abstract}
We used attitude data from the Mars Ingenuity helicopter with a simple steady-state model to estimate windspeeds and directions at altitudes of 3 meters up to 24 meters, the first time winds at such altitudes have been probed on Mars. We compared our estimates to concurrent wind data at 1.5 m height from the meteorology package MEDA onboard the Mars 2020 Perseverance rover and to predictions from meteorological models. Wind directions inferred from the Ingenuity data agreed to within uncertainties with the directions measured by MEDA, when the latter were available, but deviated from model-predicted directions by as much as $180^\circ$ in some cases. Also, the inferred windspeeds are often much higher than expected. For example, meteorological predictions tailored to the time and location of Ingenuity's 59th flight suggest Ingenuity should not have seen windspeeds above about $15\,{\rm m\ s^{-1}}$, but we inferred speeds reaching nearly $25\,{\rm m\ s^{-1}}$. By contrast, the 61st flight was at a similar time and season and showed weaker winds then the 59th flight, suggesting winds shaped by transient phenomena. For flights during which we have MEDA data to compare to, inferred windspeeds imply friction velocities exceeding $1\,{\rm m\ s^{-1}}$ and roughness lengths of more than $10\,{\rm cm}$ based on a boundary layer model that incorporates convective instability, which seem implausibly large. These results suggest Ingenuity was probing winds sensitive to aerodynamic conditions hundreds of meters upwind instead of the conditions very near Mars 2020, but they may also reflect a need for updated boundary layer wind models. An improved model for Ingenuity's aerodynamic response that includes the effects of transient winds may also modify our results. In any case, the work here provides a foundation for exploration of planetary boundary layers using drones and suggests important future avenues for research and development.
\end{abstract}



\section{Introduction} \label{sec:Introduction}
Having completed more than seventy successful flights, the Mars 2020 Ingenuity helicopter has transformed the aspiration of aerial exploration of alien worlds into reality \citep{2022cosp...44..361S}. Although intended as a technology demonstration \citep{Balaram2021, lorenz2022planetary}, Ingenuity contributed to the Mars 2020 mission by providing detailed, high resolution scouting of potential paths for the rover \citep{2023LPICo2806.2453G}. Ingenuity also collected images that allow for detection of aeolian-driven surface changes \citep{2024LPICo3040.2594G}. The pending launch and deployment of the Dragonfly rotorcraft on Saturn's moon Titan portends even more auspicious possibilities for drone-based science \citep{turtle2024dragonfly, lorenz2018dragonfly}. The combination of Titan's dense atmosphere and low surface gravity make that world a much more forgiving aerodynamic environment than Mars. As such, the Dragonfly rotorcraft will carry a rich suite of instruments, bringing the craft's mass to several hundred kg, and Dragonfly will be able to fly tens to hundreds of kilometers during its mission to explore the Selk crater region \citep{2021PSJ.....2...24L}. By contrast, the Ingenuity helicopter carries no scientific instrumentation, has a mass less than 2 kg, and can only fly a few hundred meters at a stretch \citep{9843813}. Better equipped aircraft may be possible on Mars, but even with a specialized entry-descent-and-landing approach to maximize feasible payload, Mars rotorcraft will likely be limited to only a few kg \citep{2020arXiv201006630D}, in part due to limitations of motor cooling in the thin Mars atmosphere \citep{lorenz2022planetary}. 

These limitations for Mars aerial exploration mean any way to reduce payload without reducing scientific output would be advantageous, and one obvious avenue is using the drone itself as an environmental probe. Motivated by these considerations, \citet{2022RNAAS...6..264J} recently explored using a drone to measure the near-surface wind profile, i.e. windspeed as a function of altitude. As a proof-of-concept, this effort followed on considerable previous work \citep{NEUMANN2015300, 2019Senso..19.2179B, 2022Atmos..13..551M} that showed the tilt of a stably hovering drone can scale with windspeed -- since a rotorcraft generates forward thrust, in part, by tilting into the thrust direction, the rotorcraft would have to tilt more into a stronger headwind. Drone attitude, including yaw, pitch, and roll, must be recorded for successful navigation on Mars anyway, and so these data could be a way of retrieving the near-surface wind vector without requiring additional instrumentation. 

The martian near-surface winds drive aeolian processes, which themselves impact martian geology, climate, and atmospheric evolution \citep{2001SSRv...96..393G, 2022SciA....8N3783N}, not to mention mission operations -- especially, entry, descent, and landing -- and potential future human exploration. However, the in-situ wind measurements available for Mars are relatively few and limited to single point measurements at the surface, usually measured at 1-2 m altitude \citep{1978GeoRL...5..715R, 1979JGR....84.2821R, 2000JGR...10524547S, 2010JGRE..115.0E18H, 2017Icar..291..203N, 2019Icar..319..909V, 2019Icar..319..645V, 2020NatGe..13..190B, 2020JGRE..12506493V, 2022JGRE..12707522V, 2022JGRE..12707523V, 2022SciA....8N3783N, Jiang2023}. Thus, extending wind measurements, not just across the surface of Mars but also up into the lower atmosphere, has the potential to profoundly expand our understanding of the martian environment. 

Though lacking scientific instruments, Ingenuity is equipped with a variety of engineering sensors, including a downward-facing navigation camera, lidar-based altimeter, and several accelerometers, in addition to an onboard computer -- all designed to allow the helicopter to navigate autonomously, subject to pre-designed flight operations and waypoints \citep{doi:10.2514/6.2019-1289, 9843813, Balaram2021}. The data collected by these sensors (after being processed on-board Ingenuity) were returned to Earth via Mars 2020 to allow forward planning and diagnosis of issues. As we show in this study, these same data can be used to estimate the near-surface wind vector on Mars. 

During its flights, Ingenuity probed winds between 3 and 24-meters altitude, the first time winds at such altitudes have been estimated on Mars (although some coarse constraints during the brief moments of descent and landing have been determined, \citealp[e.g.,][]{2024Icar..41516045P}). Information on the conditions at these altitudes, and especially the winds, can elucidate surface-atmosphere interactions on Mars, which are key for understanding martian climate, the water and dust cycles, and aeolian activity closer to the surface, among other features \citep{2017acm..book....1Z, 2017acm..book..106R, 2000JGR...10524547S}. In particular, under certain conditions \citep{1993BoLMe..63..323W}, the increase in windspeeds with altitude near the surface constrain the wind shear, and the wind shear controls when and whether aeolian sediments are mobilized. Indeed, the conditions under which such aeolian transport takes place remain highly uncertain for Mars \citep{NEWMAN2022637}.

The wind patterns at Jezero crater observed by Mars 2020 presented minor sol-to-sol variations generally overwhelmed by diurnal variations, in accordance with observations by previous missions on Mars outside the dust storm season. No significant variability was observed in the general shape of the diurnal cycle during the observation period \citep{2022JGRE..12707522V}. Mean wind speeds at 1.5 m altitude were $\left( 3.2 \pm 2.3 \right)\, {\rm m\ s^{-1}}$ in northern spring and summer, with 99\% of wind speeds below $10\, {\rm m\ s^{-1}}$. During the afternoon, wind peaked and reached $\left( 6.1 \pm 2.2 \right)\, {\rm m\ s^{-1}}$ \citep{2022JGRE..12707523V}. A great influence of turbulence and wave and vortex activity was observed in the wind speed variations, thus driving the highest wind speeds observed at Jezero. Mars 2020 MEDA wind data showed typical standard deviation of $\left( 0.57 \pm 0.29 \right)\, {\rm m\ s^{-1}}$ during nighttime and $\left( 1.85 \pm 0.57 \right)\, {\rm m\ s^{-1}}$ during the daytime in a ten-minute timescale, with peak values greater than $\sim 3.5\, {\rm m\ s^{-1}}$ during the daytime. These fluctuations dramatically disturbed the wind directions as well \citep{2022JGRE..12707523V}.

In this study, we present an analysis of engineering data from the Ingenuity helicopter collected during flights 1 through 5 and during flights 59 and 61, and using these data, we are able to infer windspeeds and directions. In some cases, comparable data were available from the Mars 2020 Mars Environmental Dynamics Analyzer MEDA instrument suite \citep{2021SSRv..217...48R, 2023NatGe..16...19R}, but in others we compared the winds inferred from Ingenuity's telemetry collected above 3 m altitude to model predictions. In Section \ref{sec:Description of Data}, we give a brief description of the MEDA data we used, followed by a lengthy description of the data and flight conditions for Ingenuity. In Section \ref{sec:Helicopter Dynamics Model}, we describe Ingenuity's flight dynamics and the assumptions we made to convert its attitude telemetry into wind vectors. In Section \ref{sec:Results}, we present our results, including near-surface wind profiles. In Section \ref{sec:Discussion and Conclusions}, we summarize our analysis, discuss caveats and limitations, and outline areas for future work.

\section{Description of Data}
\label{sec:Description of Data}

\subsection{Mars 2020 Wind Data}
During several of the flights we analyzed, Mars 2020's MEDA meteorological suite collected wind data to which we could compare wind derived from Ingenuity's telemetry, and so we here describe the MEDA data. The MEDA winds measured throughout the mission until the end of the nominal operations of the MEDA wind sensor were presented in \citet{2022JGRE..12707522V} and \citet{2022JGRE..12707523V}. \citet{2021SSRv..217...48R} provided a detailed description of the MEDA instrument suite. The Mars Environmental Dynamics Analyzer MEDA collects a variety of meteorological data as time series, usually at a sampling rate of 1 Hz, including pressure, relative humidity, air and ground temperature, and downwelling and upwelling, solar and infrared fluxes \citep{2023NatGe..16...19R}. Most relevant here, MEDA includes a wind sensor (WS) that consists of two booms placed at about 1.5 m above the base of the rover wheels and rotated an azimuth of 120 degree from each other around the rover mast. The horizontal component of the wind is determined with an accuracy of $1\,{\rm m\ s^{-1}}$ from 0 to $10\,{\rm m\ s^{-1}}$ and 10\% above $10\,{\rm m\ s^{-1}}$, and with an accuracy in the wind direction of $15^\circ$. MEDA wind data up to mission sol 315 are publicly available at the NASA Planetary Data System, within the derived data products (DER files). 

Normal operations for MEDA began on mission sol 15 ($L_{\rm s} \sim 12^\circ$). These operations consist of windows of time 1 hour and 5 minutes long beginning at odd Local Mean Solar Times (LMST) hours on odd sols and on even LMST hours on even sols. This approach captures both the beginning of each hour on every sol and a full hour every two sols. Availability of additional data volume and power allow additional and extended monitoring periods, during which MEDA sensors sample at 1 Hz. A higher frequency of 2 Hz has been used occasionally by some sensors (e.g., the Air Temperature Sensor ATS) to characterize turbulent phenomena.

As described in \citet{2022JGRE..12707522V} and \citet{2022GeoRL..4900126L}, the MEDA wind sensor suffered damage during the onset of the MY36/2022A regional dust storm that prevented the current engineering retrieval from deriving wind magnitudes. It is expected that wind data will be available when the retrieval algorithm for each boom is modified to focus on the non-damaged boards of the sensor, although greater uncertainty in the wind data will be involved in the retrievals. Therefore, no wind data are available after sol 315 (Table 1 in \citealt{2022JGRE..12707522V}). However, wind data are available during flights 1-5. At that period, the Mars 2020 rover was within about $100\, {\rm m}$ distance of the helicopter. Data from MEDA during flights 59 and 61 are not publicly available, and, in any case, the rover was more than $300\,{\rm m}$ distant from the helicopter. Thus, we assume the MEDA wind data collected during flights 1-5 can be reasonably compared to those inferred from Ingenuity's flight data, as described below. Of course, because Mars 2020 was $\sim$100 m distant from Ingenuity during flights 1-5, we may expect that the wind fields seen by Ingenuity and by Mars 2020 do not exactly match up either spatially or temporally. However, we leave a detailed exploration of this issue to future work. 

We also note that, in addition to the dedicated wind speed sensors on MEDA, some constraints on wind and turbulence have been derived from measurements by the microphone on the SuperCAM instrument, \citep[e.g.,][]{2023JGRE..12807547S}. In particular, sound recordings were made during a number of Ingenuity flights when the helicopter was close enough to the rover for its sounds to be detected \citep{lorenz2023sounds}. Of the flights analyzed in the present paper, flights 4 and 5 were recorded, and wind noise was particularly strong at the beginning of flight 4. 

In any case, the high level of turbulence intensity observed by Mars 2020 \citep{2022JGRE..12707523V, 2023JGRE..12807607P, 2023JGRE..12807547S} could provoke discrepancies between Mars 2020 and Ingenuity locations, both in wind speeds and directions. The wind conditions were also highly variable during Ingenuity's first flights, as reported in Table 1 of \citet{2022JGRE..12707522V}, for periods spanning $\pm$5 min around the Ingenuity flight phase. The mean wind speed during the flights was $4.6\, {\rm m\ s^{-1}}$ at 1.5 m.

\subsection{Ingenuity Flight Data}\label{sec:Description_of_Flights_and_Flight_Data}
In this section, we describe the locations and times for each flight analyzed. The first five flights were part of the prime mission of the technology demonstration. These flights progressed from simple up and down flights to lateral flights up to $\sim$75 m. The first four flights landed at the same airfield near the landing site with the rover fairly close by in order to image Ingenuity during flight \citep{2022cosp...44..362G}. After the first five technology demonstration flights, the helicopter entered an operational extended mission to scout out ahead of the rover and to support other engineering and science investigations \citep{2023LPICo2806.2453G, 2024LPICo3040.2594G}. \citet{2022JGRE..12707605L} also used images of dust stirred up by Ingenuity while in flight to constrain aeolian thresholds, and \citet{lorenz2023sounds} explored the sound of Ingenuity's flights propagating through the Martian atmosphere, which is strongly attenuating of higher acoustic frequencies. 

\begin{figure}
    \centering
    \includegraphics[width=0.5\linewidth]{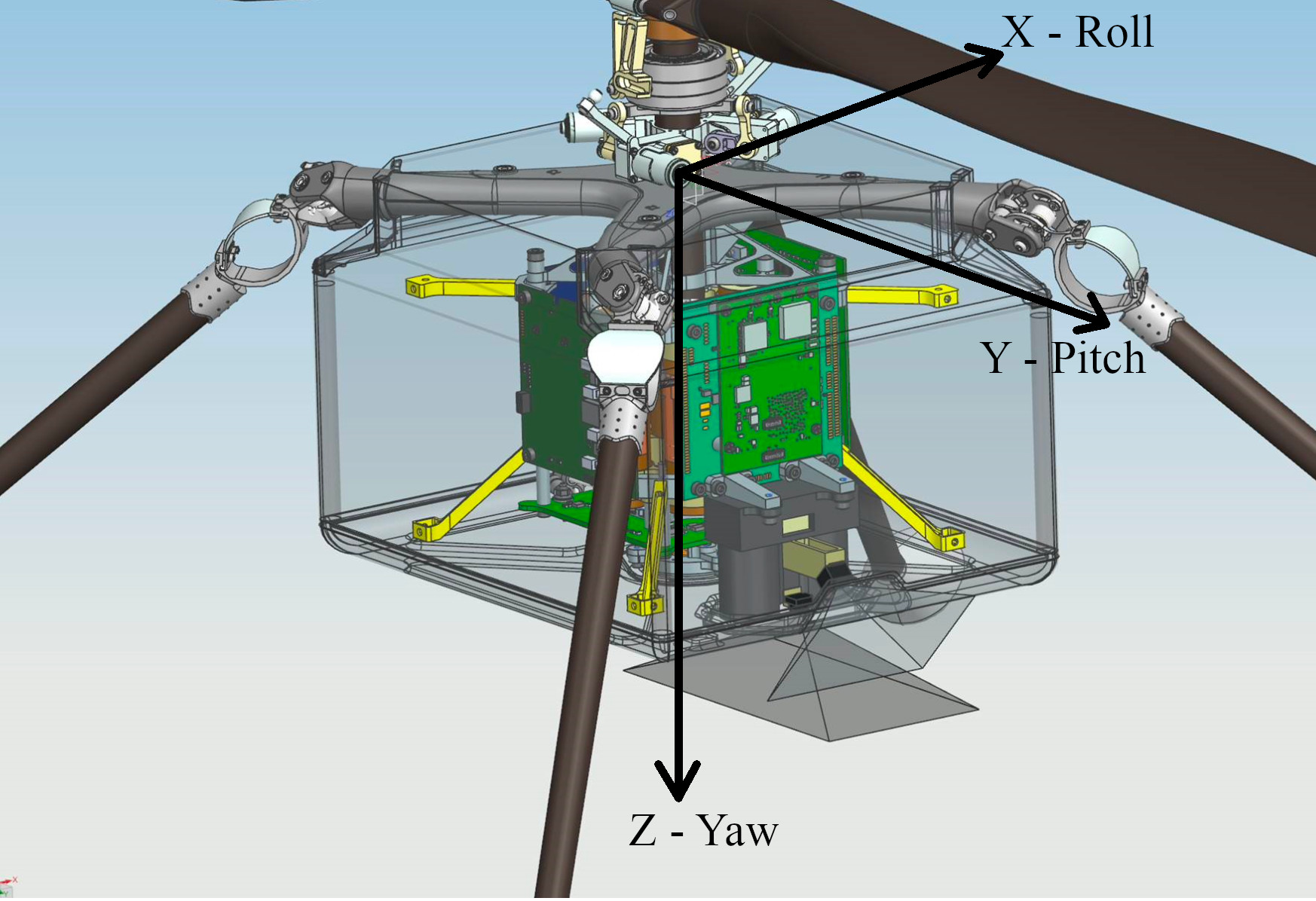}
    \caption{Zoomed-in view of Ingenuity's fuselage and interior, as adapted from the Mars 2020 Software Interface Specification. The helicopter camera locations and viewing frustrums are shown. The yaw ($z$), pitch ($y$), and roll ($x$) axes are also shown. For example, a given pitch angle corresponds to a right-handed rotation about the $y$ axis.}
    \label{fig:Mars2020_SIS}
\end{figure}

The Ingenuity telemetry we analyzed are returned at 500 Hz and are derived from a combination of onboard sensors and visual velocimetry, as described in Section \ref{sec:Helicopter Dynamics Model}. The telemetry data are defined as follows: 
\begin{itemize}
    \item altitude - Ingenuity's height above the local surface;
    \item $dx/dt$ and $dy/dt$ - Ingenuity's groundspeed as measured along the $x$ and $y$ axes (shown in Figure \ref{fig:Mars2020_SIS}) affixed to the helicopter's body;
    \item yaw - the angle between the north cardinal direction and Ingenuity's $x$ axis;
    \item pitch - the angle for rotation about Ingenuity's $y$ axis;
    \item roll - the angle for rotation about Ingenuity's $x$ axis;
    \item tilt - the angle between the local vertical and Ingenuity's propeller stalk as determined from the pitch and roll angles (see Equation \ref{eqn:tilt_angle});
    \item tilt azimuth - the angle between the north cardinal direction and the projection of Ingenuity's propeller stalk on the ground.    
\end{itemize}

In the next section, we also describe the meteorological conditions during each flight as inferred from MEDA data, when available. The time stamps between the telemetry and MEDA data streams are synchronized to about 2.5-second precision. Given that we employ averages over 10 seconds or longer for our analysis here, that synchronization is sufficient for our purposes. See Table 1 for a summary. The average MEDA wind data used for each Ingenuity flight (when available) are also included for reference. The environmental conditions around each flight are also presented in Table 1 in \citet{2022JGRE..12707522V}.

\subsection{Conditions for Each Flight}
\subsubsection{Flight 1}

\begin{figure}
    \centering
    \includegraphics[width=\textwidth]{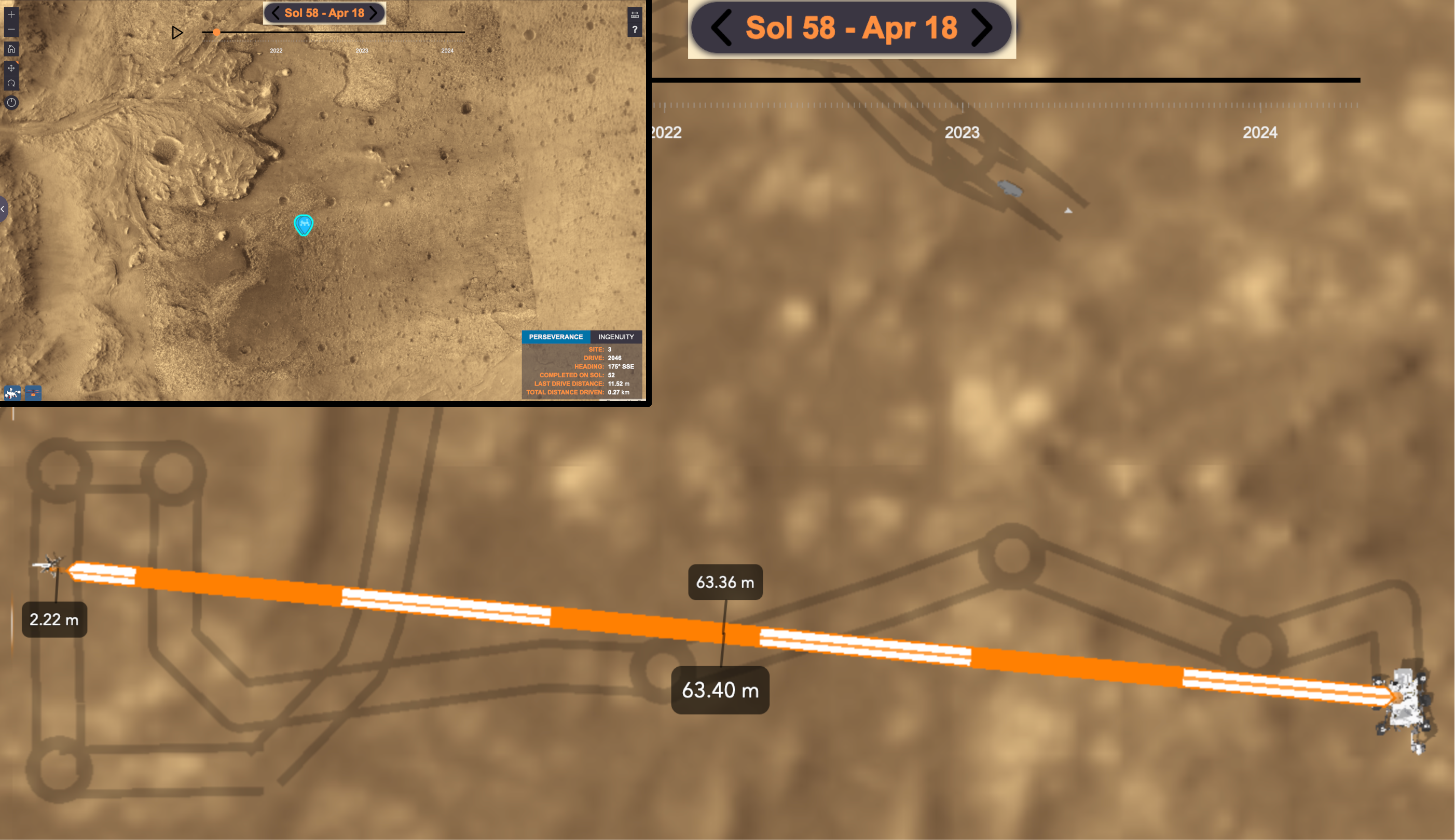}
    \caption{Location of flight 1 with an inset at the upper left showing the larger context. For flight 1, Ingenuity hovered at a lateral distance due west from Perseverance of about 63 m. North is up in both the detail and inset panels. All geography figures come from the Mars 2020 Mission Tracker website -- \url{https://experience.arcgis.com/experience/84a507c94b3a4b32b654e7f544ad1d52}.}
    \label{fig:Flight_1_Geography}
\end{figure}

\begin{figure}
    \centering
    \includegraphics[width=\textwidth]{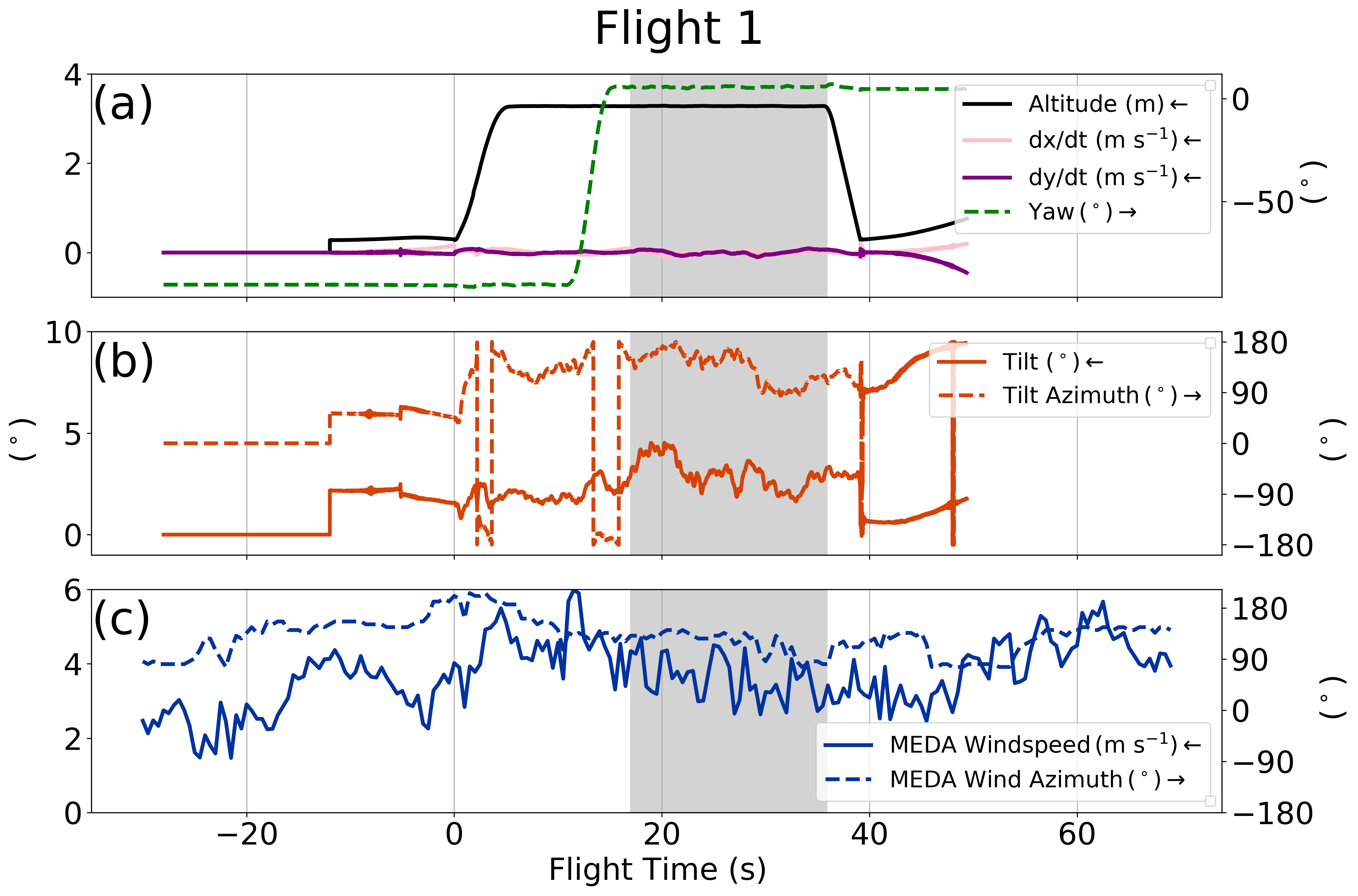}
    \caption{Telemetry streams from Ingenuity (panels a and b) and MEDA wind data (panel c) for flight 1. The legends provide the curve labels. Solid lines can be read off the left y-axis (as also indicated by the small arrows in the legend), while dashed lines can be read off the right y-axis. (Note that an azimuth of $-180^\circ$ is equivalent to $+180^\circ$, and lines are not wrapped around in this figure to improve visibility.) In (a), the solid black line is Ingenuity's altitude in meters, the pink line is $dx/dt$ which is the component of Ingenuity's groundspeed measured along the $x$ axis, the purple line is $dy/dt$ which is the component of Ingenuity's groundspeed measured along the $y$ axis, and the dashed green line is Ingenuity's yaw in degrees. In (b), the solid orange line is the tilt angle $t$, and the dashed orange line is the tilt azimuth. In (c), the solid blue line is the windspeed measured by MEDA, and the dashed blue line is the wind azimuth measured by MEDA. The shaded grey region shows the time window during which we assumed the drone was in trimmed flight conditions (i.e., no net accelerations or torques), and it is during these periods that we estimated windspeeds.}
    \label{fig:Flight_1_telemetry_and_winds}
\end{figure}


Flight 1 took place on mission sol 58 (2021 Apr 18) on $L_{\rm s} = 33.67268^\circ$ at about 12:13 LTST/12:30 LMST at $18.44486^\circ$ N, $77.45102^\circ$ E. Figure \ref{fig:Flight_1_Geography} shows the geography, and Figure \ref{fig:Flight_1_telemetry_and_winds} shows the telemetry. During flight 1, Ingenuity was about 63 m due west of Perseverance, and the flight involved a 40-sec hover at 3.25 m altitude. Figure \ref{fig:Flight_1_telemetry_and_winds} also shows the windspeed and azimuth (the direction from which the winds blew) measured by the MEDA wind sensors during the flight. Winds seen by MEDA averaged $3.5\,{\rm m\ s^{-1}}$ during the hover and came from about $126^\circ$ azimuth. For winds at 1.5 m height, these are typical values for midday in the first 250 sols of the mission \citep{2023NatGe..16...19R}.

\subsubsection{Flight 2}

\begin{figure}
    \centering
    \includegraphics[width=\textwidth]{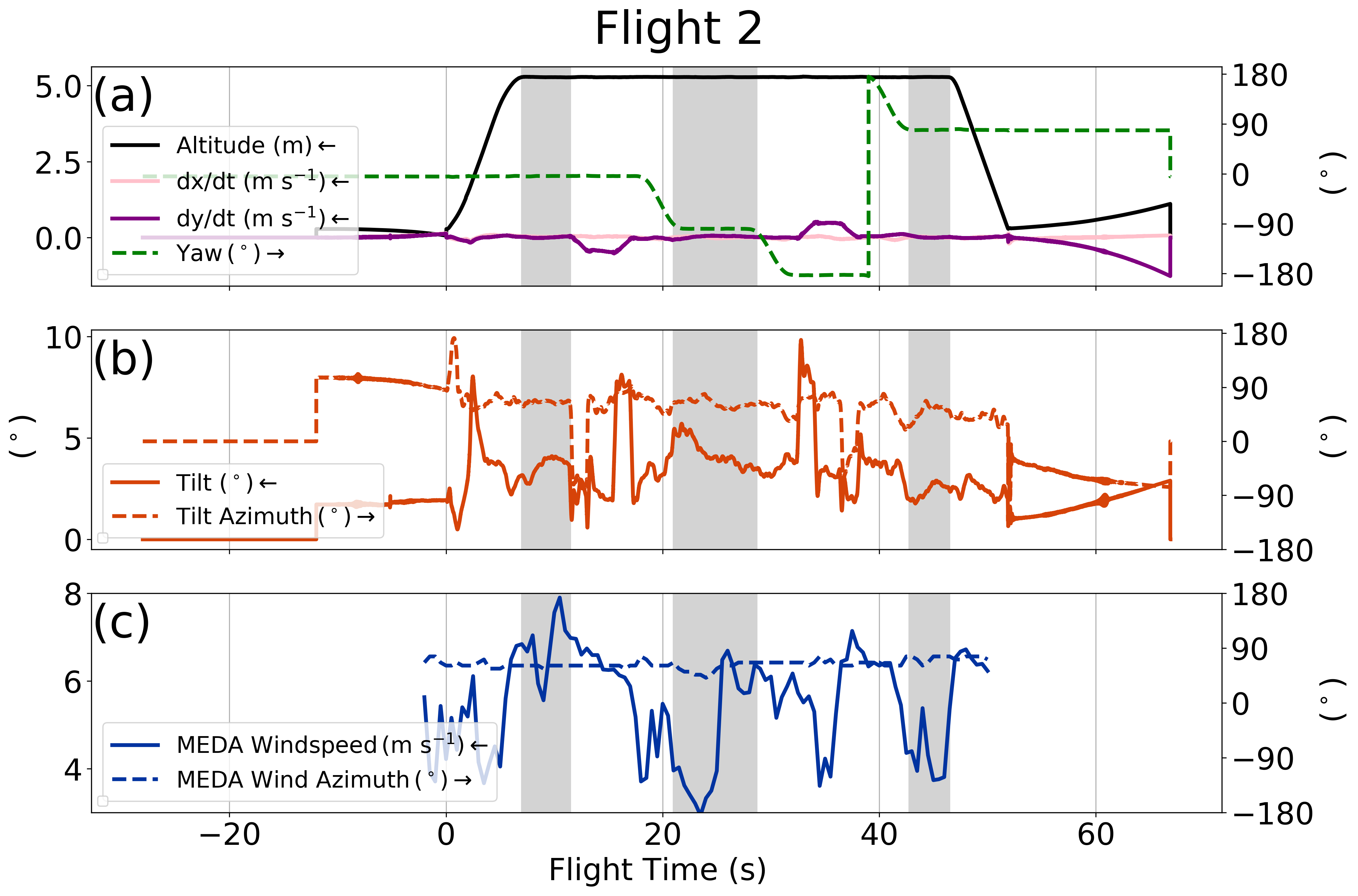}
    \caption{Telemetry streams (panels a and b) and MEDA wind data (panel c) for flight 2. The legends provide the curve labels, and line colors and styles and shaded grey regions have the same meanings as in Figure \ref{fig:Flight_1_telemetry_and_winds}. (Note that an azimuth of $-180^\circ$ is equivalent to $+180^\circ$.)}
    \label{fig:Flight_2_telemetry_and_winds}
\end{figure}

Flight 2 took place on mission sol 61 (2021 Apr 22) on $L_{\rm s} = 35.07378^\circ$ at about 12:14 LTST/12:30 LMST at $18.44486^\circ$ N, $77.45102^\circ$ E. The flight involved a 51.9-sec, 2-m out-and-back trip at 5.28-m altitude from nearly the same take-off point as for flight 1. Winds seen by MEDA averaged about $6\,{\rm m\ s^{-1}}$ during the flight from about $60^\circ$ azimuth. Figure \ref{fig:Flight_2_telemetry_and_winds} shows the telemetry and MEDA wind data.

\subsubsection{Flight 3}
\begin{figure}
    \centering
    \includegraphics[width=\textwidth]{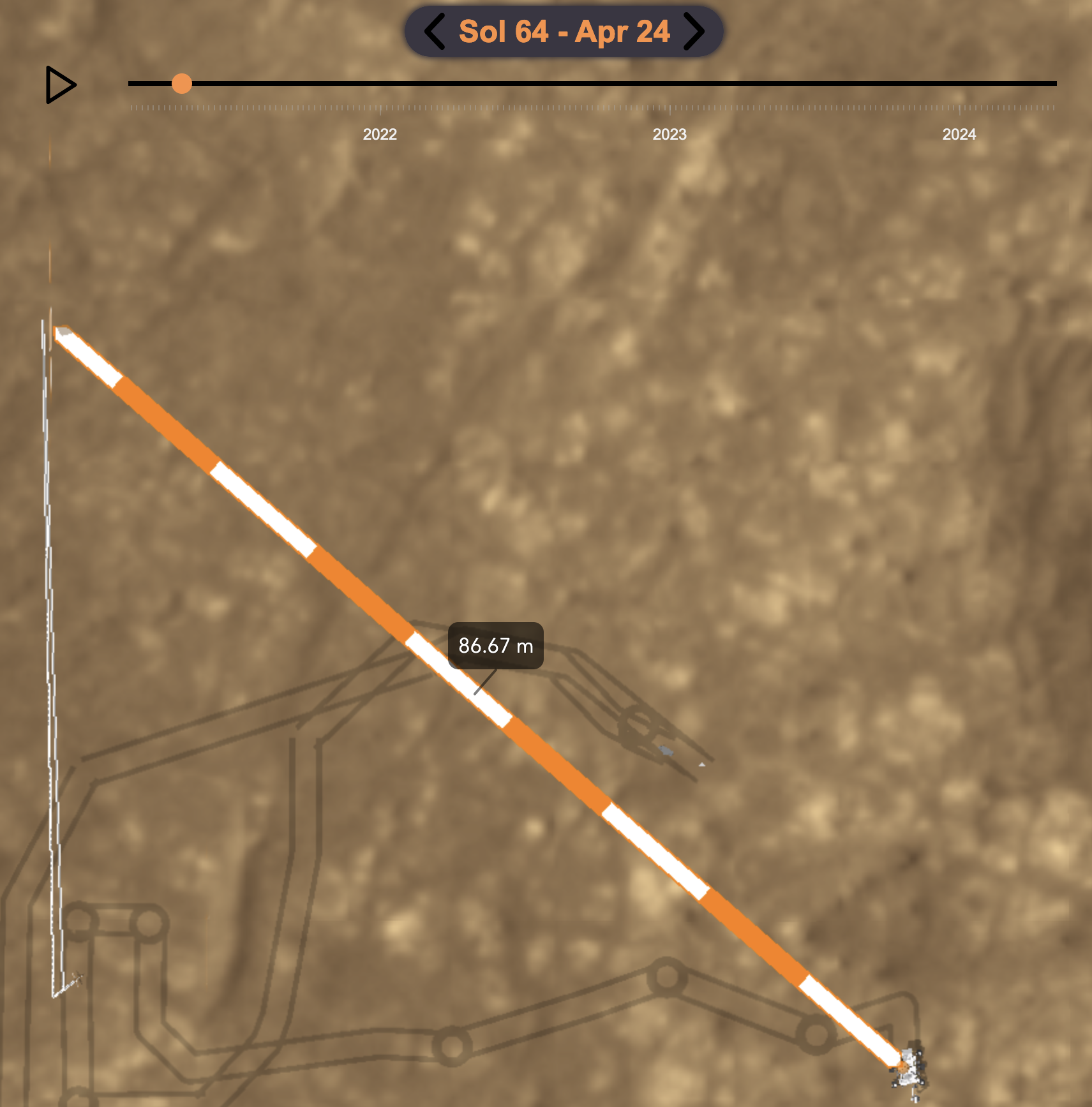}
    \caption{Geography of flight 3. Ingenuity flew about 50 meters due north at an altitude of 5.28 m and then flew due south, returning to its take-off point very near to the take-off point for flights 1 and 2. The lateral distance indicated by the ruler in the figure is the distance along the ground.}
    \label{fig:Flight_3_Geography}
\end{figure}

\begin{figure}
    \centering
    \includegraphics[width=\textwidth]{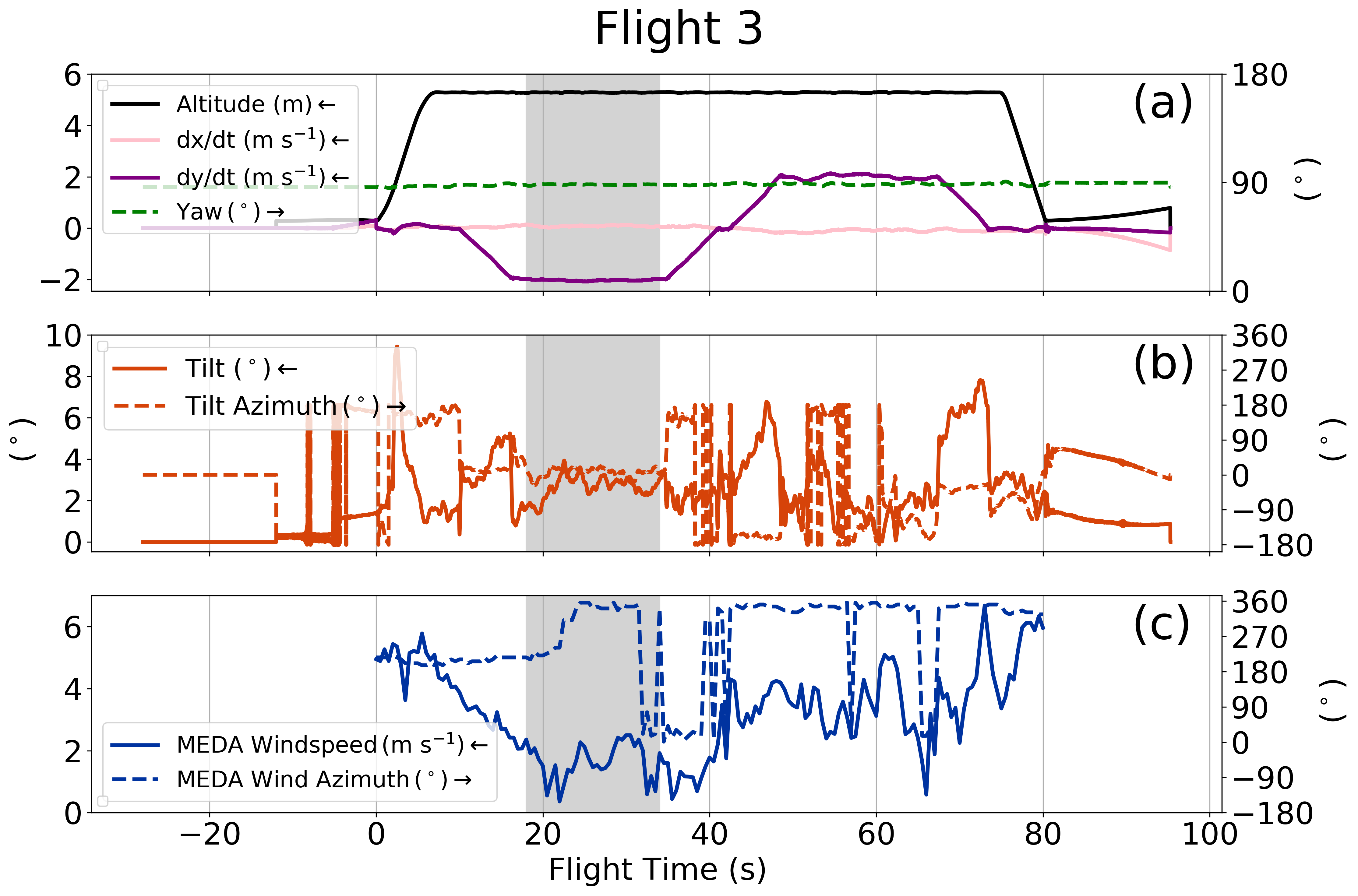}
    \caption{Telemetry streams (panels a and b) and MEDA wind data (panel c) for flight 3. The legends provide the curve labels, and line colors and styles and shaded grey regions have the same meanings as in Figure \ref{fig:Flight_1_telemetry_and_winds}. (An azimuth of $360^\circ$ is equivalent to an azimuth of $0^\circ$, and an azimuth of $-180^\circ$ is equivalent to $+180^\circ$.)}
    \label{fig:Flight_3_telemetry_and_winds}
\end{figure}

Flight 3 took place on mission sol 64 (2021 Apr 25) on $L_{\rm s} = 36.47081^\circ$ at about 12:15 LTST/12:30 LMST at $18.44486^\circ$ N, $77.45101^\circ$ E and lasted 80.3 seconds. During flight 3, Ingenuity took off from a point about 63 m due west of Perseverance and flew about 50 m to the north at a groundspeed of about $2\,{\rm m\ s^{-1}}$ and an altitude of 5.28 m and then returned to its take-off point. The take-off point was nearly the same point as for flights 1 and 2. Figure \ref{fig:Flight_3_Geography} show the geography, and Figure \ref{fig:Flight_3_telemetry_and_winds} shows the telemetry and MEDA wind data. Prior to flight time 20 seconds, MEDA measured winds blowing in from an azimuth of about $200^\circ$ (Figure \ref{fig:Flight_3_telemetry_and_winds}) and then blew in from about $340^\circ$ starting at flight time 27 seconds with some variability thereafter. Windspeeds seen by MEDA during flight 3 varied from about $2\,{\rm m\ s^{-1}}$ to $6\,{\rm m\ s^{-1}}$ but were much more stable during the flight phase we analyzed (shown in grey in Figure \ref{fig:Flight_3_telemetry_and_winds}).

\subsubsection{Flight 4}
\begin{figure}
    \centering
    \includegraphics[width=\textwidth]{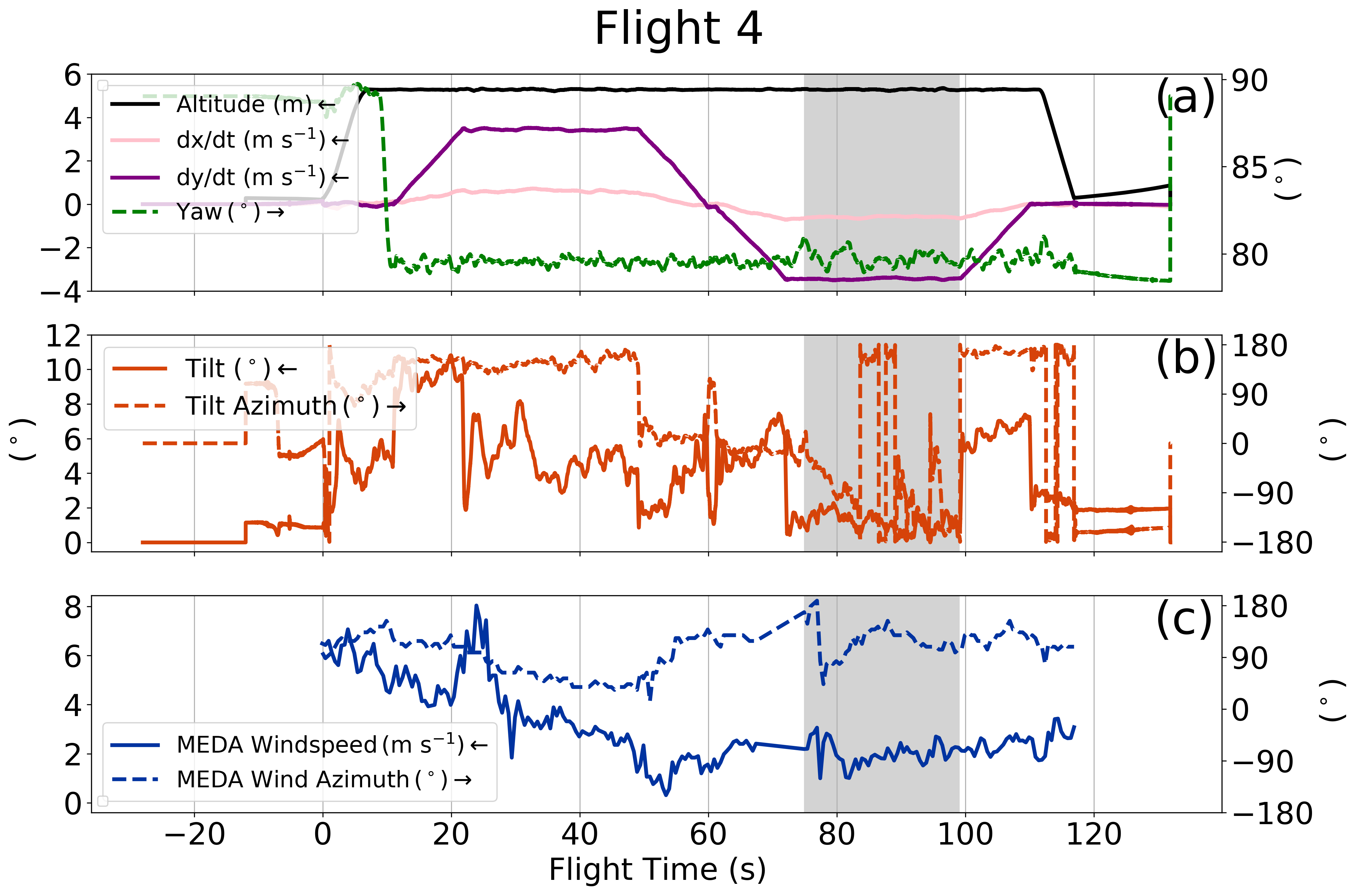}
    \caption{Telemetry streams (panels a and b) and MEDA wind data (panel c) for Flight 4. The legends provide the curve labels, and line colors and styles and shaded grey regions have the same meanings as in Figure \ref{fig:Flight_1_telemetry_and_winds}. (An azimuth of $-180^\circ$ is equivalent to $+180^\circ$.)}
    \label{fig:Flight_4_telemetry_and_winds}
\end{figure}

Flight 4 took place on mission sol 69 (2021 Apr 30) on $L_{\rm s} = 38.79059^\circ$ at about 12:17 LTST/12:30 LMST at $18.44486^\circ$ N, $77.45112^\circ$ E. The flight involved a 117-sec, 135-m out-and-back trip at 5.28 m altitude toward the south at about $3.5\,{\rm m\ s^{-1}}$ groundspeed from nearly the same take-off point as the prior flights. Winds seen by MEDA varied between around $2.0\,{\rm m\ s^{-1}}$ from $115^\circ$ during the flight. Figure \ref{fig:Flight_4_telemetry_and_winds} shows the telemetry and MEDA wind data. \citet{lorenz2023sounds} indicates that wind noise was particularly prominent from 0-5s, 15-25s, and 45-55s into the flight, somewhat consistent with the MEDA data. While the last of these acoustic gusts does appear to manifest in the helicopter tilt data, it is not evident in the MEDA timeseries. The $dy/dt$ ground velocity (purple line in Figure \ref{fig:Flight_4_telemetry_and_winds}) was detected by the microphone as a Doppler shift of the 84 Hz blade passage frequency. 

\subsubsection{Flight 5}

\begin{figure}
    \centering
    \includegraphics[width=\textwidth]{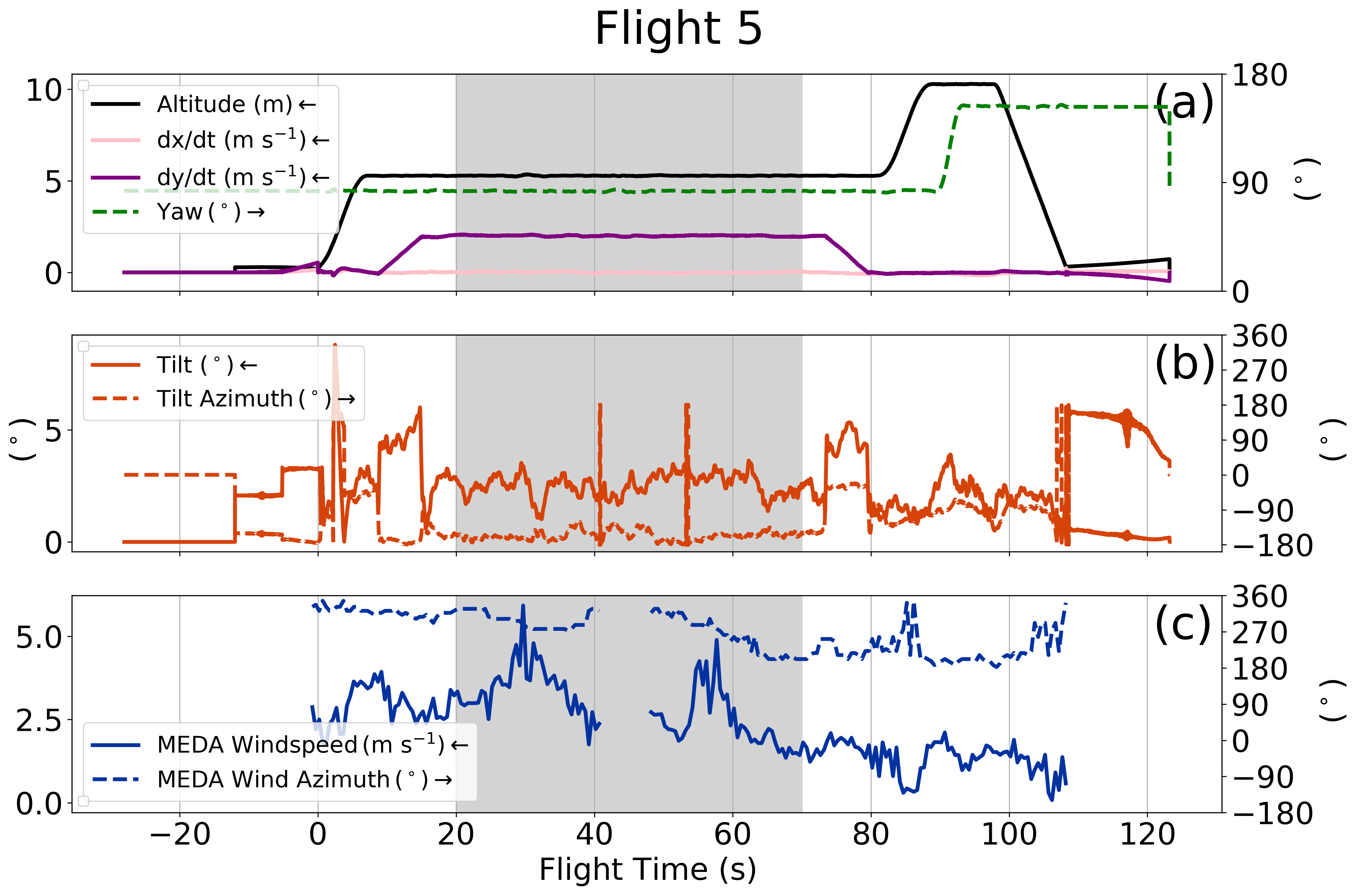}
    \caption{Telemetry streams (panels a and b) and MEDA wind data (panel c) for Flight 5. The legends provide the curve labels, and line colors and styles and shaded grey regions have the same meanings as in Figure \ref{fig:Flight_1_telemetry_and_winds} (An azimuth of $360^\circ$ is equivalent to an azimuth of $0^\circ$ and $-180^\circ$ is equivalent to $+180^\circ$.) }
    \label{fig:Flight_5_telemetry_and_winds}
\end{figure}

Flight 5 took place on mission sol 76 (2021 May 7) on $L_{\rm s} = 42.02140^\circ$ at about 12:19 LTST/12:30 LMST at $18.44267^\circ$ N, $77.45139^\circ$ E. The flight involved a 108-sec, 130-m trip at 5.28 m altitude toward the south at about $2\,{\rm m\ s^{-1}}$ groundspeed from the landing point for flight 4, putting the landing point about 110 m toward the southwest from Perseverance. Just prior to landing, the drone rotated about $50^\circ$ in yaw and ascended to about 10 m altitude. Winds seen by MEDA averaged $2.9\,{\rm m\ s^{-1}}$ from about $290^\circ$ azimuth. Figure \ref{fig:Flight_5_telemetry_and_winds} shows the telemetry and MEDA wind data. The tilt data show prominent excursions around 12 and 75 s, which are related to maneuvering, and brief wind gusts are evident in the SuperCAM microphone record at 5, 25, 40 and 75 seconds (see Figure 7 of \citealt{lorenz2023sounds}).

\subsubsection{Flight 59}
\begin{figure}
    \centering
    \includegraphics[width=\textwidth]{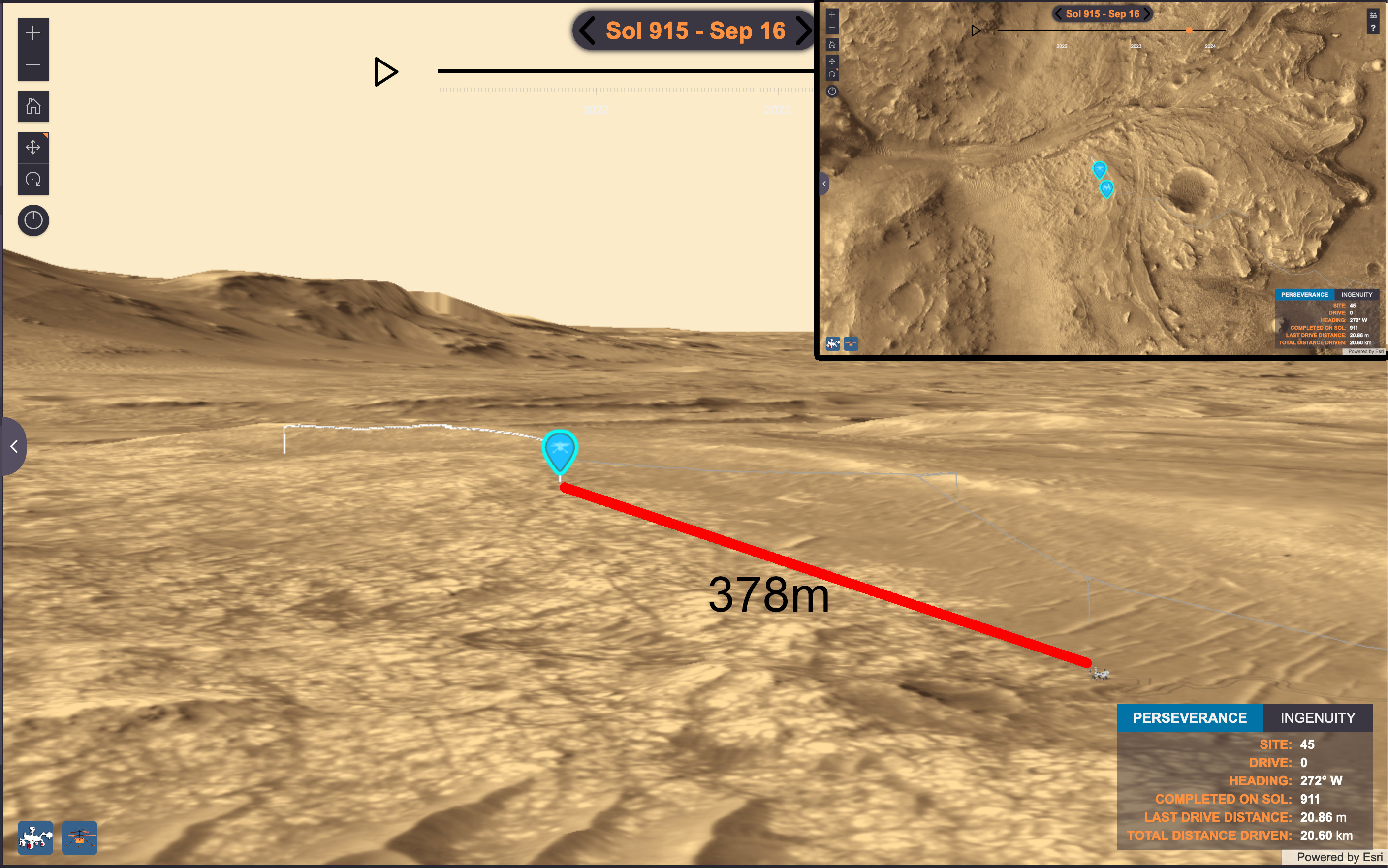}
    \caption{Location of flight 59 with an inset at the upper right showing the larger context, with north pointing forward into the distance along the surface in oblique panel and up in the inset. For flight 59, Ingenuity took off from a point northwest from Perseverance at a distance of about 378 m. The long white strand in the larger, oblique view shows the flight path toward the take-off point for flight 61.}
    \label{fig:Flight_59_Geography}
\end{figure}

\begin{figure}
    \centering
    \includegraphics[width=0.8\textwidth]{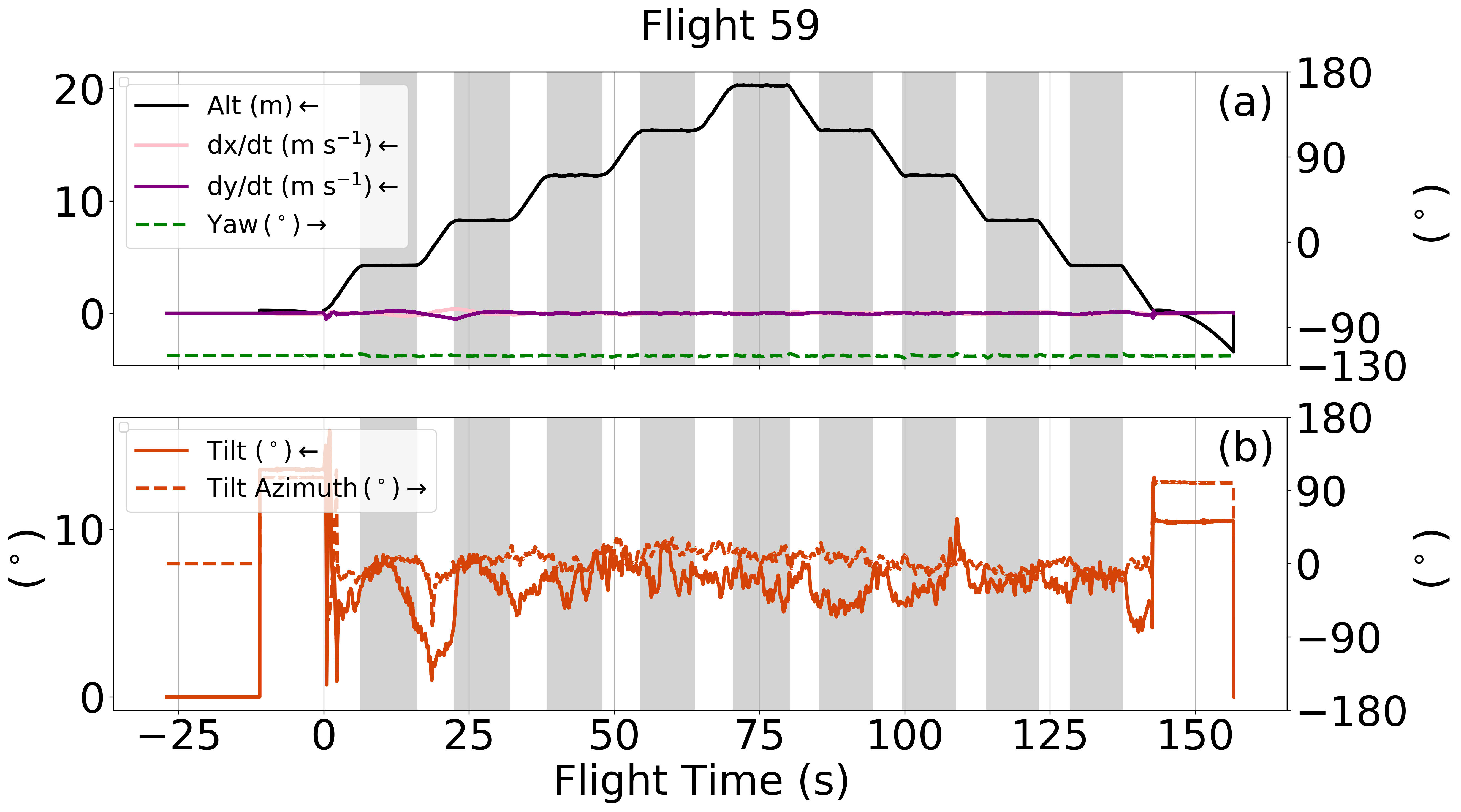}
    \caption{Telemetry streams for flight 59. The legend provides the curve labels, and line colors and styles and shaded grey regions have the same meanings as in Figure \ref{fig:Flight_1_telemetry_and_winds}. There are no MEDA wind data against which to compare.}
    \label{fig:Flight_59_Telemetry}
\end{figure}

Flight 59 took place on mission sol 915 (2023 Sep 16) on $L_{\rm s} = 108.2954^\circ$ at about 11:25 LTST/11:02 LMST at $18.489398^\circ$ N, $77.348344^\circ$ E. During flight 59, Ingenuity took off from a mega-ripple measuring about 1 m in height and 6 m in width on the delta within Jezero about 378 m northwest of Perseverance and completed a stair-step ascent with 10-sec hovers at five different altitudes -- 4.25, 8.25, 12.25, 16.25, and 20.25 m -- and then hovered at each of those altitudes on the descent, with the goal of profiling the winds at altitude. Figure \ref{fig:Flight_59_Geography} shows the geography, and Figure \ref{fig:Flight_59_Telemetry} shows the telemetry. There are no MEDA wind data available for this flight; however, according to some models, this is the period of the year with the most intense winds at Jezero \citep{2021SSRv..217...20N}.

\subsubsection{Flight 61}
\begin{figure}
    \centering
    \includegraphics[width=0.8\textwidth]{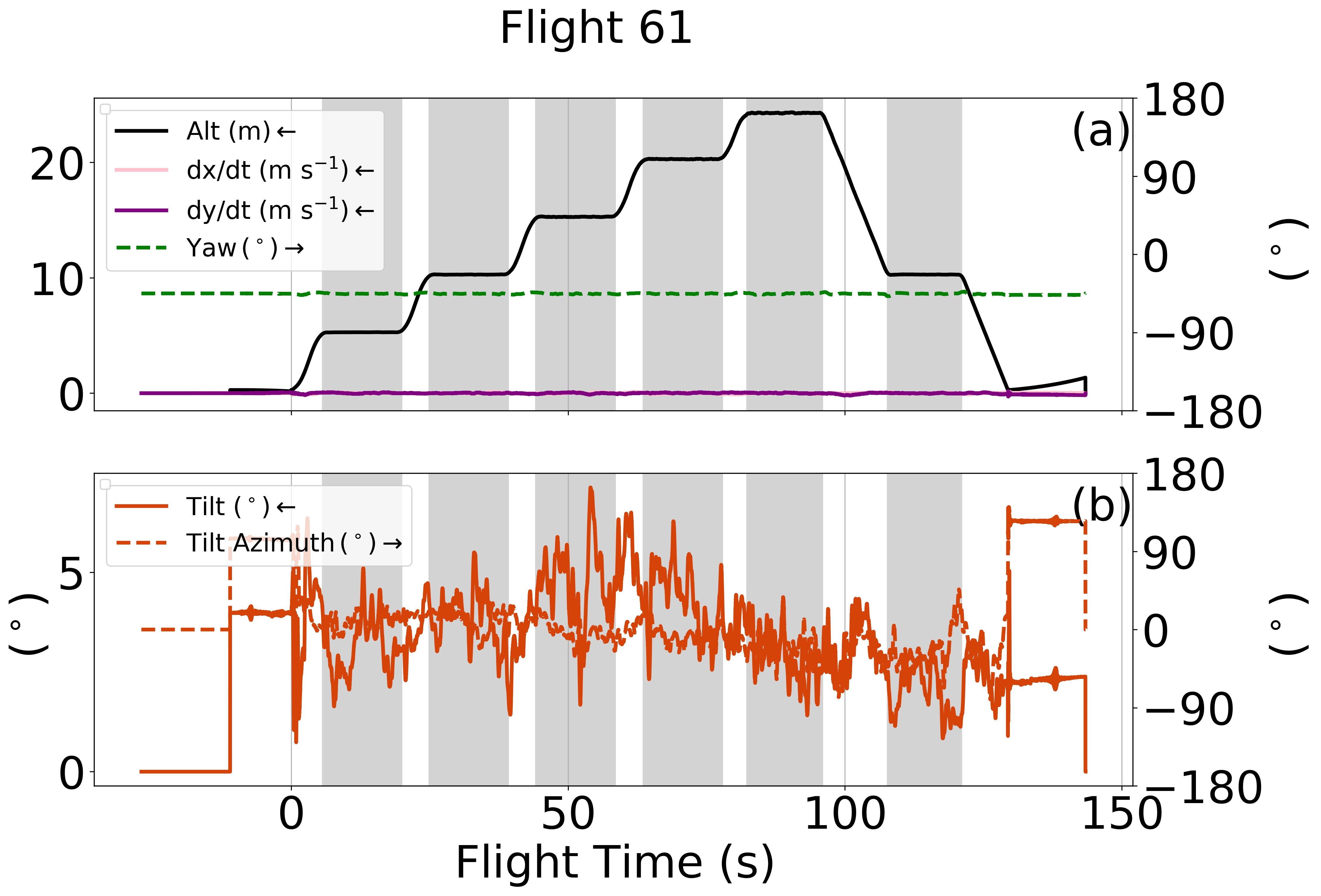}
    \caption{Telemetry streams for flight 61. The legend provides the curve labels, and line colors and styles and shaded grey regions have the same meanings as in Figure \ref{fig:Flight_1_telemetry_and_winds}. There are no MEDA wind data against which to compare.}
    \label{fig:Flight_61_Telemetry}
\end{figure}

Flight 61 took place on mission sol 933 (2023 Oct 5) on $L_{\rm s} = 116.5606^\circ$ at about 12:28 LTST/12:02 LMST at $18.493916^\circ$ N, $77.345112^\circ$ E, from an inter-ripple plain about 325 m northwest of the take-off point for flight 59 and about 336 m northwest of Perseverance (also on the delta). Like flight 59, this flight also involved a stair-step ascent, this time with 14-sec hovers at five different altitudes -- 5.28, 10.28, 15.28, 20.28, and 24.28 m -- and then a 14-sec hover at 10.28 m altitude during the descent. Figure \ref{fig:Flight_61_Telemetry} shows the telemetry. There are no MEDA wind data available for this flight.

\section{Helicopter Dynamics}\label{sec:Helicopter Dynamics Model}
Ingenuity's attitude is determined from a combination of onboard sensors, digital elevation models (DEMs), and visual velocimetry, and \citet{doi:10.2514/6.2019-1289} provided a detailed overview of Ingenuity's flight control system. Ingenuity collects images from the Return to Earth (RTE) color camera, which points at $45^\circ$ from the helicopter body, and the Navigation camera (Navcam), which is nadir-pointed. The Navcam is a 640 × 480 grayscale camera, collecting images at 30 Hz. These images provide the basis for visual navigation via the specially developed algorithm, Minimal Augmented state algorithm for Vision-based Navigation MAVeN \citep{doi:10.2514/6.2019-1411}. MAVeN, implemented as an Extended Kalman Filter, tracks visual features on the martian surface and continually updates a state vector for the helicopter, which includes components for position, velocity, attitude, and biases for the onboard accelerometer and gyro. 

\citet{doi:10.2514/6.2019-1411} described a comprehensive suite of experiments, including field deployments of an engineering model of Ingenuity, to assess the accuracy and biases of MAVeN navigation. Tests showed that the estimates of position and velocity only drifted away from their true values by $\leq 1\, {\rm m}$ and $\leq 40\,{\rm cm\ s^{-1}}$, respectively, after 200 seconds of flight (longer than the flights analyzed here). Roll, pitch, and yaw estimates from MAVeN also accumulated error over time, drifting away from the true values by about $1.5^\circ$ but only after 120 seconds of flight. We see no evidence for these kinds of drift in our data, though, and, in any case, only the very last windspeed data point from flight 59 (Figure \ref{fig:Flight_59_Telemetry}) derives from flight times after 120 seconds.

Before that drift, the roll, pitch, and yaw estimates exhibited smooth though small excursions around the true values, with timescales of 2-3 seconds. Kalman filter solutions are known to exhibit such correlated noise \citep[e.g.,][]{9521680}, and those excursions are apparent in the Ingenuity tilt data. The Ingenuity data from Mars also exhibit smaller and shorter-term ($\sim$ 0.5 seconds) oscillations on top of such excursions. These oscillations may likewise arise from the Kalman filter solution, or they may represent the response of Ingenuity to eddies. (The field tests from \citealt{doi:10.2514/6.2019-1411} did not assess the impacts of wind or eddies.) Indeed, the MEDA wind data show $\sim$1-${\rm m\ s^{-1}}$ variations in windspeed with a similar timescale. Whatever their origin, we employ an approach that accommodates these oscillations -- see Section \ref{sec:Results}.

\begin{figure}
    \centering
    \includegraphics[width=0.8\textwidth]{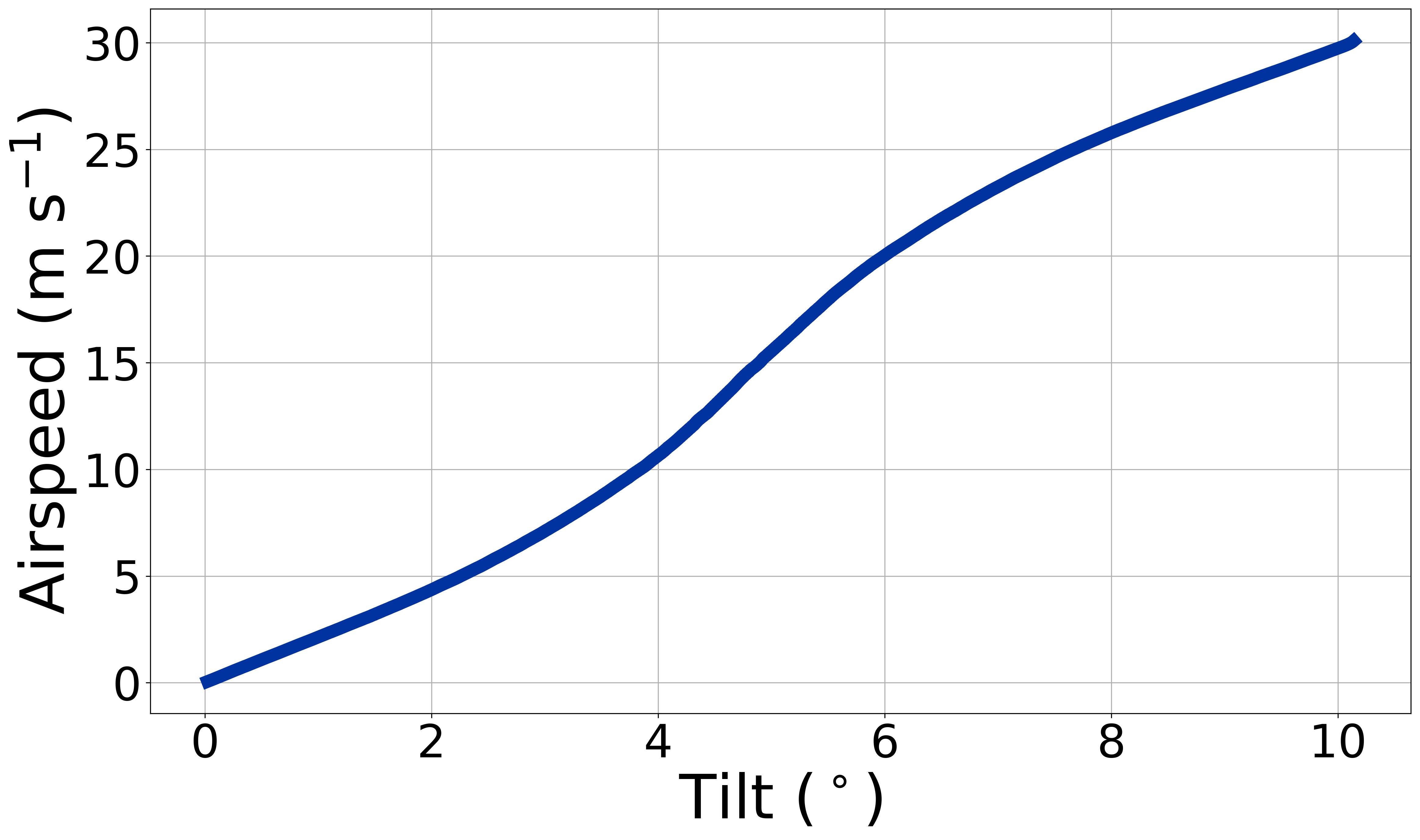}
    \caption{Based on \citet{Dull2022}, our model for relating the drone's tilt angle (given in terms of pitch and roll by Equation \ref{eqn:tilt_angle}) to airspeed.}
    \label{fig:Dull2022_Fig13}
\end{figure}
In order to translate Ingenuity's attitude telemetry into a wind vector, we need to understand its response to a given wind vector. For our analysis here, we employed an approach based on simplified modeling of the Ingenuity helicopter. More complex and comprehensive models are currently not available, but for the first-order analysis here, our approach suffices, though future work should expand on our results. 

To estimate the horizontal wind velocity vector $\vec{\upsilon}_{\rm Ingenuity}$ from Ingenuity's telemetry, we need an estimate of the horizontal airspeed vector $\vec{\upsilon}_{\rm air}$ and horizontal groundspeed velocity vector $\vec{\upsilon}_{\rm ground}$. In principle, $\vec{\upsilon}_{\rm ground} = \vec{\upsilon}_{\rm Ingenuity} + \vec{\upsilon}_{\rm air}$. The groundspeed vector comes directly from the $dx/dt$ and $dy/dt$ velocities reported in the telemetry data (pink and purple lines in the telemetry figures). These velocities are converted into north- and eastward ground speeds using the yaw angle, shown as a dashed green line in the telemetry figures. For estimating the airspeed vector, we considered only portions of the flights when we expect the helicopter is in a trimmed flight condition, i.e., it experienced no net torques or accelerations. In practice, we selected periods when the accelerations were inferred to be below a few tens of ${\rm cm\ s^{-2}}$ since there are no periods when accelerations were identically measured to be zero. We also focused on times when the MEDA-measured windspeed was not varying significantly, i.e., when the windspeed changed by only about $\pm 1\,{\rm m\ s^{-1}}$ and the azimuth was relatively steady over the whole analyzed phase. Of course, the winds measured at MEDA do not directly reflect the winds seen by Ingenuity, but this requirement provides a useful gauge of the expected gustiness at Ingenuity. This approach required some subjective selection of the averaging windows, but via experiment, we found that expanding or contracting the chosen averaging windows slightly did not significantly impact our results.

Turning back to consideration of trimmed flight conditions, an example of such a state is a stable, fixed-point hover. Trimmed conditions occurred during certain portions of all the flights analyzed here. The averaging windows we used are shown as shaded regions in the telemetry figures. During these trimmed flight phases, the various forces acting on the helicopter -- air drag, weight, etc. -- are nearly balanced by the helicopter's thrust. \citet{Dull2022} outlined the results of several computational fluid dynamics CFD simulations and experiments in JPL's Space Simulator facility with an Ingenuity engineering model to explore the power and thrust characteristics of Ingenuity under martian conditions for a nominal propeller rotation speed of 2,600 RPM but various collective and differential orientations. Among those results, \citet{Dull2022} reported the expected airspeed of the helicopter during trimmed forward flight, shown in Figure \ref{fig:Dull2022_Fig13} which is adapted from that study. This key result allows us to translate the helicopter's attitude into an airspeed vector.

Next, we describe the key attitude elements. Ingenuity's pitch axis (the $y$-axis) points along the direction into which the RTE camera faces, the yaw axis (the $z$-axis) points downward toward the ground, and the roll axis (the $x$-axis) points at a right angle to both. The angle between the vertical  and the propeller stalk, we call the tilt angle $t$, and it is given in terms of the pitch $p$ and roll $r$ angles by 
\begin{equation}
    \cos t = \cos p \cos r.\label{eqn:tilt_angle}
\end{equation}

We also calculated what we call the tilt azimuth, i.e., the orientation of Ingenuity's negative $z$-axis relative to north. We calculated the tilt azimuth by first applying the yaw, pitch, and roll rotation matrices sequentially and in that order to a unit vector affixed to Ingenuity and pointing along its negative z-axis (i.e., upward along the propeller stem). Then we calculated the projection of the resulting unit vector along the north direction. We use the tilt azimuth to determine airspeed azimuth and therefrom the wind azimuth. The tilt azimuth should point in the cardinal direction of Ingenuity's thrust vector. If, for example, Ingenuity were hovering stably, in-place against a headwind, the thrust vector would point into the direction from which the wind were coming and would be larger for faster windspeeds. Consequently, we can use the magnitude of the tilt angle to estimate airspeed and tilt azimuth to estimate airspeed direction. 

In summary, the wind azimuth inferred this way typically matches the wind azimuth measured by MEDA to within a few degrees when the drone hovers in-place and less well (tens of degrees) when the drone has a substantially non-zero groundspeed. This mismatch is likely due not to a mismatch between the actual tilt azimuth and airspeed direction but rather to uncertainties on inferring the tilt azimuth from the available roll, pitch, and yaw data. Consider, for instance, the case that the tilt angle is very nearly zero. Then, the tilt azimuth itself could be very uncertain and could easily swing between different values. Indeed, the window of time during flight 4 telemetry that we analyzed shows significant variations in tilt azimuth because the tilt angle itself is only about $1^\circ$. In addition, the agreement in wind direction at the rover and Ingenuity suggest that both experience eddies rolling over them in the same background wind direction, with the eddies perhaps perturbing the wind speed more than direction. 

It is also important to note that the CFD model of Ingenuity used in \citet{Dull2022} incorporates Ingenuity's coaxial rotors, rotor mast, airframe, legs, and solar array and assumes forward thrust pointing in the direction of the y axis. In other words, in the model, the roll angle $r$ is held fixed at $0^\circ$, and forward motion (with respect to the air) is modeled as the result of rotating the helicopter into the direction of the motion. The relationship reported in \citet{Dull2022} is, strictly, the airspeed vs.~pitch and not vs.~tilt. During the flights we analyze, Ingenuity rotates about both the roll and pitch axes to achieve thrust. However, the dominant aerodynamic elements for Ingenuity are its 1.5-m diameter propellers (as compared to the 136 mm × 195 mm × 163 mm fuselage). The propellers would be expected to exhibit a symmetric aerodynamic influence, and therefore the thrust should depend mostly on the tilt angle. Throughout the analysis described below, several checks bolster this assumption. We leave to future work the expected small corrections from fuselage and solar panel drag and assume a direct relationship between airspeed and tilt, as illustrated in Figure \ref{fig:Dull2022_Fig13}.

Another important assumption we make is that, after maneuvering into a given attitude, Ingenuity quickly settles into a trimmed flight condition with respect to the air drag. In other words, we assume Ingenuity exhibits no long-term hysteresis and that its attitude in a trimmed condition, averaged over several seconds, directly reflects the aerodynamic conditions averaged over the same window. We conservatively averaged over time windows that lasted 10 seconds or more and that start at least one second after the last maneuver. For example, during flight 1, a rotation in yaw was executed between flight times 11 and 16 seconds (solid, green line in Figure \ref{fig:Flight_1_telemetry_and_winds}), and so we started averaging up tilt values after flight time of 17 seconds (shaded grey region in Figure \ref{fig:Flight_1_telemetry_and_winds}). This assumption is justified on the basis of flight video and telemetry. Mars 2020 SuperCam collected video of flights 1 and 59, which show no evidence for dynamical instability or undamped wobble, suggesting that the helicopter achieved trimmed flight conditions very quickly after settling at an altitude. 

\section{Results}\label{sec:Results}

\include{Windspeeds_table}

\begin{figure}
    \centering
    \includegraphics[width=\textwidth]{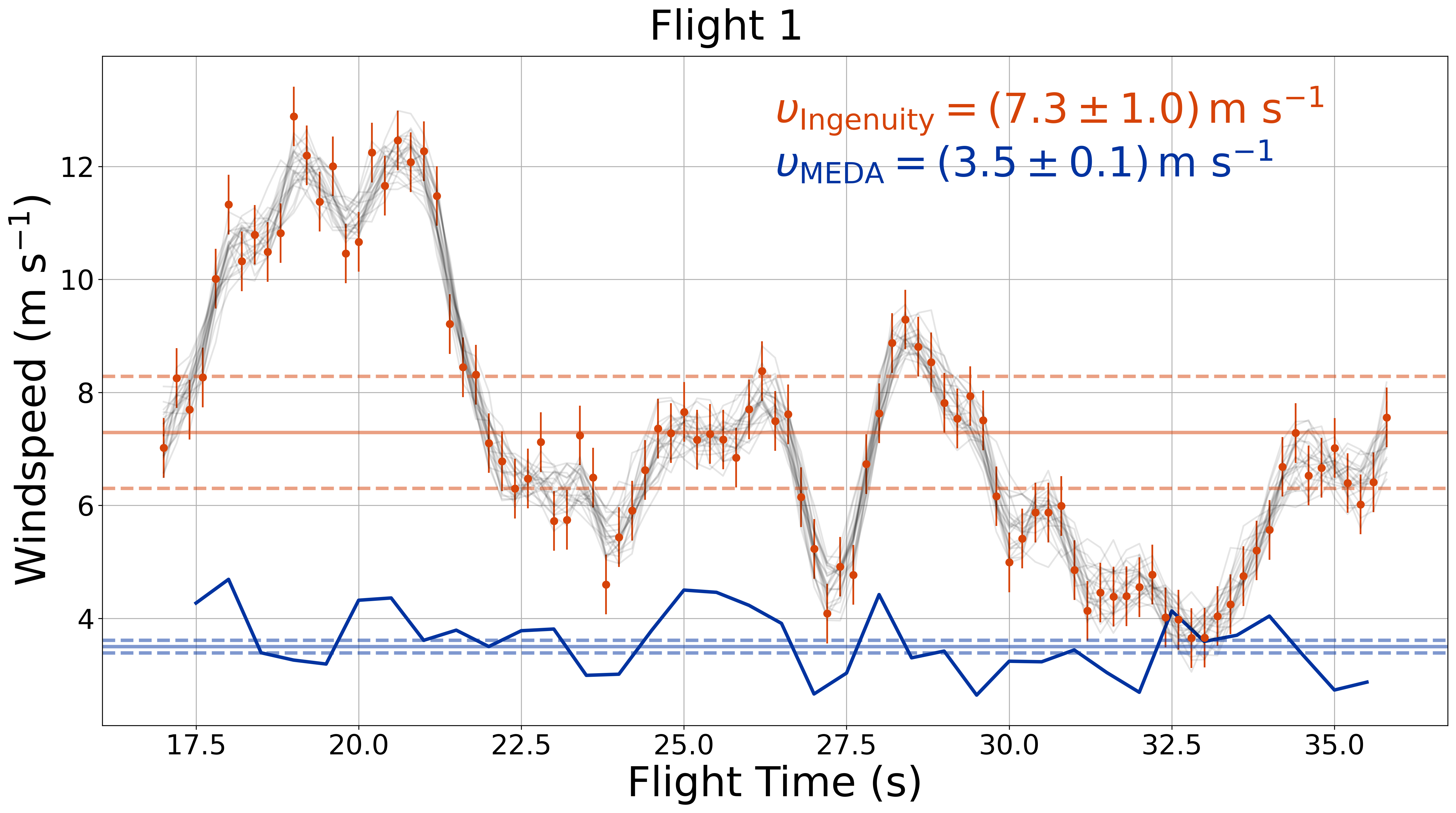}
    \caption{Time series of windspeeds during flight 1. The orange dots represent the windspeeds inferred from Ingenuity's tilt, downsampled by a factor of one hundred. The grey lines show the Gaussian process model, and the resulting average and uncertainty is shown in orange at the top right (also illustrated by the horizontal orange lines). The MEDA windspeed is shown in blue.}
    \label{fig:Flight_1_Windspeeds}
\end{figure}

Figure \ref{fig:Flight_1_Windspeeds} shows how we developed our estimate of average windspeed during flight 1. Using the tilt-airspeed relation shown in Figure \ref{fig:Dull2022_Fig13}, we converted the point-to-point tilt telemetry to windspeed as a function of time. (For comparison, the concurrent MEDA windspeed data are shown in blue.) As discussed above, Ingenuity's telemetry time series show correlated, i.e., red, noise. Such noise has the tendency to skew model fit parameters in ways that are not well described by a Gaussian, i.e., white, noise model \citep{2006gpml.book.....R}. By contrast, the MEDA windspeed data during each flight time window we analyzed (the grey, shaded periods in the telemetry figures) are well described by a white noise model, with an average value over the flight 1 time window of $\upsilon_{\rm MEDA} = 3.5\,{\rm m\ s^{-1}}$ and a standard error $\sigma_{\upsilon_{\rm MEDA}} = 0.1\,{\rm m\ s^{-1}}$. (We checked the assumption of a white noise model for the MEDA data by comparing that approach to the results of the Gaussian process noise model described next.) To account for this more complex noise, we applied a Gaussian process model to the flight 1 windspeed data, which allows for a non-zero covariance between data points (i.e., the scatter in one data point impacts scatter of nearby points). To describe the covariance, we used a Matern-3/2 kernel (as implemented in the george package -- \citealp{2015ITPAM..38..252A}), and we employed an MCMC analysis (as implemented in the emcee package -- \citealp{2013PASP..125..306F}) to estimate the posterior distribution for the average windspeed. (Because the Gaussian process analysis is computationally intensive, we reduced the telemetry sampling rate by a factor of 100 to render the problem tractable.) For the MCMC analyis, we incorporated 32 walkers and two burn-in phases of a few thousand links each, before the final production chain of a few thousand more links. We assessed convergence by estimating the auto-correlation time for the chains and requiring it to be more than 50 times shorter than the total chain length. Figure \ref{fig:Flight_1_Windspeeds} shows the resulting Gaussian process model for the noise as grey lines. (Since they represent a noise model, there is no \emph{one} best-fit noise, just a range of possibilities.) We take the median value from the posterior as the average and the maximum of the 16\% and 84\% percentiles as the uncertainty. The resulting average windspeed estimate for flight 1 is $\upsilon_{\rm Ingenuity} = \left( 7.3 \pm 1.0\right)\,{\rm m\ s^{-1}}$. By comparison, the standard error would have given an uncertainty on the Ingenuity windspeed of $2.4\,{\rm m\ s^{-1}}$, much larger than the Gaussian process estimate.

For flight 2, we employed the same approach, although in this case, the time span of tilts analyzed was not continuous since Ingenuity conducted several maneuvers during which our assumption of trimmed flight probably breaks down. Numerical experimentation with synthetic datasets to mimic these data indicated that averaging windspeeds over non-continuous windows of time can still return reasonable time-averages. For flight 2, we estimated $\upsilon_{\rm Ingenuity} = \left( 9.1 \pm 2.0 \right)\,{\rm m\ s^{-1}}$ and $\upsilon_{\rm MEDA} = \left( 5.8 \pm 0.1 \right)\,{\rm m\ s^{-1}}$. For flight 3, we considered only the southward portion of the flight, a 16-second window, since, during the northward portion, the MEDA winds varied considerably, as did Ingenuity's reported tilt. Of course, the large variability in tilt may reflect actual blustery conditions, but we leave consideration of such conditions to future work. For flight 3, we estimated $\upsilon_{\rm Ingenuity} = \left( 4.1 \pm 0.7 \right)\,{\rm m\ s^{-1}}$ and $\upsilon_{\rm MEDA} = \left( 1.6 \pm 0.1 \right)\,{\rm m\ s^{-1}}$. For similar reasons, we only considered the northward, 24-second portion of flight 4 when estimating windspeed since the MEDA wind and Ingenuity tilt data showed considerable variability during the southward portion. For flight 4, we estimated $\upsilon_{\rm Ingenuity} = \left( 4.1 \pm 0.7 \right)\,{\rm m\ s^{-1}}$ and $\upsilon_{\rm MEDA} = \left( 2.0 \pm 0.1 \right)\,{\rm m\ s^{-1}}$. Flight 5 had a much longer stretch of relatively constant wind and tilt data during a 50-second southward flight. For flight 5, we estimated $\upsilon_{\rm Ingenuity} = \left( 4.3 \pm 0.3 \right)\,{\rm m\ s^{-1}}$ and $\upsilon_{\rm MEDA} = \left( 2.9 \pm 0.1 \right)\,{\rm m\ s^{-1}}$. 

We employed the same approach for windspeeds estimated during flights 59 and 61; however, since those flights involved stable hovers at set altitudes, we estimated average windspeeds during each of those hovers individually. Since those flights were specially tailored for wind profiling (and flight 59 provides a serendipitous check on our approach), we start our discussion of results with those flights.

As discussed above, we also estimated wind azimuth from tilt azimuth. During each window from which we estimated windspeed, we take the median value for the inferred windspeed azimuth angle. We also estimated uncertainties on this azimuth by using the median absolute deviation. The wind azimuths inferred from Ingenuity telemetry agreed to within these formal uncertainties with the MEDA azimuths, when the latter were available. Table 1 shows all the results from these wind vector estimates.

\subsection{Flights 59 and 61}
\begin{figure}
    \centering
    \includegraphics[width=\textwidth]{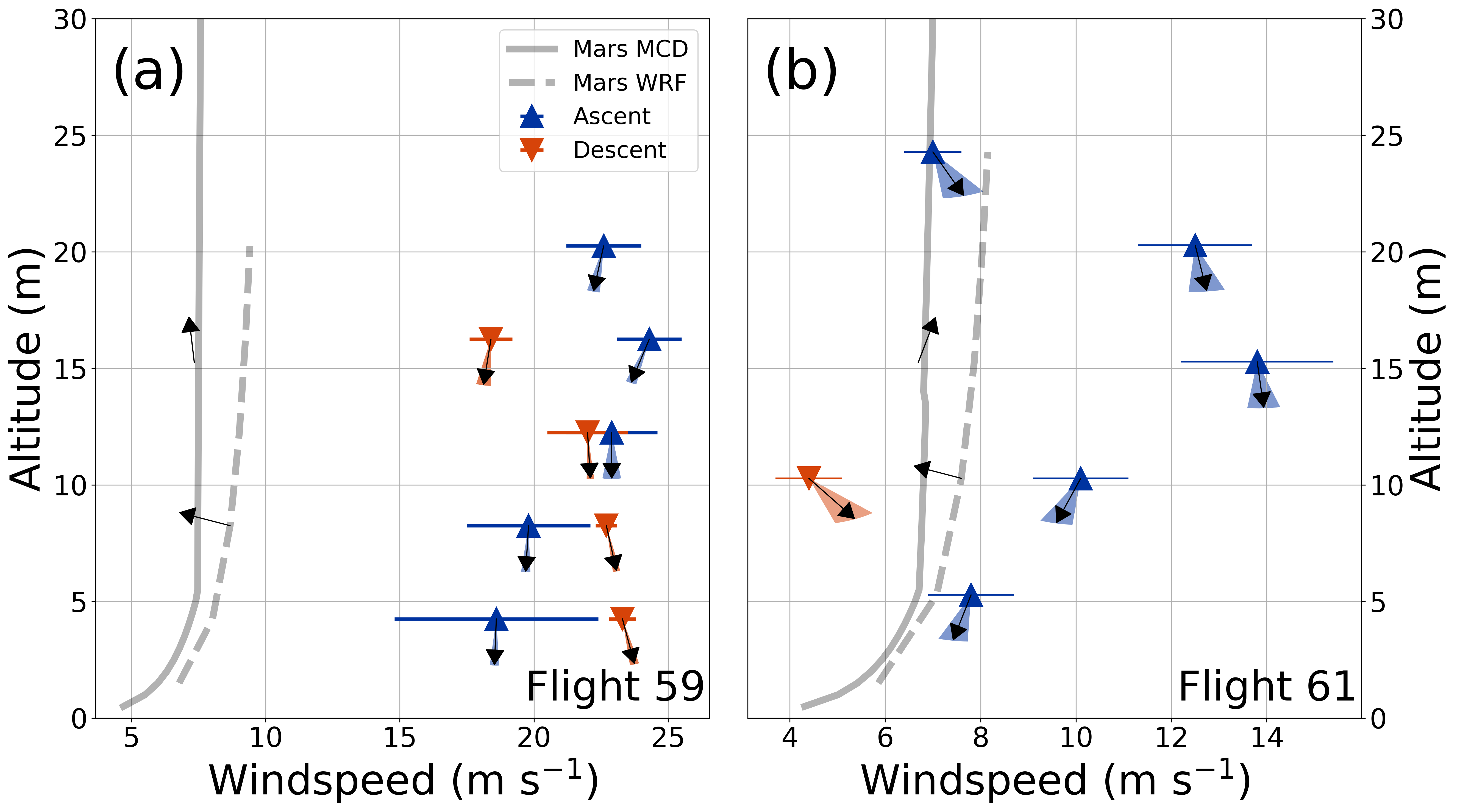}
    \caption{Windspeeds and directions inferred during flights 59 (a) and 61 (b) over the time window when Ingenuity was hovering. The colored wedges show the range of uncertainties on the wind direction for points inferred from Ingenuity telemetry (see Table 1). The blue, upward-pointing triangles show inferred values during the ascent, while the orange, downward-pointing triangles are from descent. The solid, grey lines show the corresponding windspeed profile from the Mars MCD model, and the dashed grey lines show the wind profile extrapolated from MarsWRF. The arrows attached to points and to the profiles indicate the direction \emph{into which the winds are blowing}, with north at the top of the plot.}
    \label{fig:Flights_59_and_61_Wind_Profiles}
\end{figure}

Figure \ref{fig:Flights_59_and_61_Wind_Profiles} shows the windspeeds and directions inferred from flights 59 and 61. Recall that these flights involved in-place hovers at several consecutive altitudes and that no MEDA data are available for comparison. We compared these results to predictions from two meteorological models: (1) the Mars Global Climate Database, compiled from the Laboratoire de Meteorologie Dynamique Mars Planetary Climate Model \citep{1999JGR...10424155F, 2018fmee.confE..68M}, which provides predicted windspeeds and directions as a function of altitude for a specified time of day and location. We used the downloadable, high resolution version of the database, which combines high resolution (32 pixels/degree) MOLA topography and pressure records with the MCD surface pressure in order to compute surface pressure as accurately as possible. The latter is then used to reconstruct vertical pressure levels and hence yield high resolution values of atmospheric variables; and (2) the MarsWRF model \citep{2007JGRE..112.9001R, 2012Icar..221..276T, 2017Icar..291..203N, 2019JGRE..124.3442N, 2021SSRv..217...20N}, which can simulate nested higher-resolution domains within a global context and includes the radiative effects of carbon dioxide gas and ices, aerosol dust, and water vapor and water ice, along with cycles of carbon dioxide and dust. The MarsWRF simulations included here ``nest in'' to a grid spacing of $\sim$1.4 km in the innermost nest over Jezero. Of course, comparing to model results differs from comparing to in situ measured data as the model results may be inaccurate. 

Figure \ref{fig:Flights_59_and_61_Wind_Profiles} shows the resulting wind profiles and direction averaged over the profiles (MCD and MarsWRF both predict no significant variation in azimuth over their respective profiles). For the MarsWRF wind profile, we developed model wind predictions across several sols around the flight 59 and 61 sols to explore a range of reasonable wind profile for at 1.5, 10, and 40 m altitude. To each of these numerically derived profiles, we fit a boundary layer wind profile as described in detail below (see Equation \ref{eqn:wind_profile}), calculated an average profile, and then interpolated to the specific altitudes at which Ingenuity hovered.

Starting with flight 59 (Figure \ref{fig:Flights_59_and_61_Wind_Profiles}(a)), the windspeed estimates span a nominal range from 18.6 to $24.3\,{\rm m\ s^{-1}}$ but with uncertainties as large as $3.8\,{\rm m\ s^{-1}}$ (see Table 1). Thus, windspeeds inferred from Ingenuity agree at all altitudes to within $1.5\sigma$, consistent with a flat wind profile, a result that agrees with the MCD results that predict a nearly flat wind profile above 5 m altitude. Approximating the near-surface winds with a logarithmic wind profile under neutral stability (see Section \ref{sec:Flights 1 through 5} and \citealt{1988aitb.book.....S}), the MCD-predicted friction velocity $u_\star = 0.3\,{\rm m\ s^{-1}}$ and roughness scale $z_\star = 1\,{\rm cm}$. (N.B., we use $z_\star$ instead of the usual $z_0$ for roughness length throughout this paper.) For the parameters, a logarithmic wind profile would be expected to show a difference in the windspeed between 4.25 and 20.25 m altitude of about $1.1\,{\rm m\ s^{-1}}$, which is smaller than the mutual uncertainties for the windspeeds inferred. Thus, at the relatively small altitudes probed by Ingenuity, we would not expect to see a discernible difference in windspeeds. We have the same expectations for the flight 61 results (Figure \ref{fig:Flights_59_and_61_Wind_Profiles}(b)). 

By contrast, the average friction velocity for flight 59 predicted by MarsWRF is $u_\star = 1\,{\rm m\ s^{-1}}$, which translates into a difference in windspeeds between 4.25 and 20.25 m altitude of about $4\,{\rm m\ s^{-1}}$ and is the formal difference observed for the ascent leg for the best-fit values, though the windspeeds at these two altitudes agree to within uncertainties. The descent leg saw a larger windspeed at 4.25 than at 20.25 m altitude, though. For flight 61, MarsWRF predicts $u_\star = 0.8\,{\rm m\ s^{-1}}$, which translates to a difference in windspeeds between 5.28 and 24.28 m altitude of about $3\,{\rm m\ s^{-1}}$. Unfortunately, the actual observed difference in windspeeds between these two altitudes is formally negative. Comparing, instead, windspeeds at 5.28 and 20.28 m altitude, we expect very nearly the same difference, $3\,{\rm m\ s^{-1}}$, while the observed difference is $4.7\,{\rm m\ s^{-1}}$. 

What about how the magnitudes, rather than the differences, of the predicted and inferred windspeeds themselves compare? For flight 59, the inferred windspeeds are all much larger than the speeds predicted by both models. Perhaps this mismatch suggests the tilt-windspeed relation used here (Figure \ref{fig:Dull2022_Fig13}) is inaccurate, but the telemetry from flight 59 provides a check that corroborates that relationship, as discussed next. For flight 61, too, the inferred windspeeds exceed the predicted windspeeds for all points except during the descent at 10.28 m. 

Turning to the wind direction, the inferred directions are very nearly opposite from the model predictions (although the two sets of predictions themselves disagree by several tens of degrees). This result seems to contradict extensive modeling of the regional wind flows around Jezero. \citet{2021SSRv..217...20N} compared nine sets of simulations to predict the meteorology and aeolian activity of the Mars 2020 landing site region. That study argued that flows in and out of Jezero were largely directed by local and regional scale slopes, with daytime winds tending to flow up the local slope from SE to NW (or ESE to WNW) -- this trend is reflected in the arrows attached to the MCD and Mars WRF wind profiles in Figure \ref{fig:Flights_59_and_61_Wind_Profiles}. The nighttime flows show the reverse, with downslope flows into Jezero dominating (from NW to SE or WNW to ESE). 

The topography in which flights 59 and 61 took place may play some role in locally redirecting the flows, thereby explaining the mis-match between inferred and predicted winds. Mars 2020 is surrounded by complex topography with high relief that could influence the local wind speed and direction in various ways \citep{https://doi.org/10.1029/2023EA003045}. The inset for Figure \ref{fig:Flight_59_Geography} shows several nearby features with high relief. The delta deposit scarp $\sim$1 km to the southeast has a $\sim$40 m drop to lower delta deposits, Belva crater on the delta deposit $\sim$1.5 km to the east is $\sim$60 m deeper than the delta deposit in which it impacted, the main east-flowing channel of Neretva Vallis cuts $\sim$10 m into the delta deposit $\sim$1 km to the north, and $\sim$4.5 km to the west Neretva Vallis cuts through the western Jezero crater rim  which lies $\sim$650 m above the elevation of Mars 2020's Flights 59 and 61. More locally, Ingenuity and Perseverance were located on fractured Noachian terrain $\sim$100 m west of a $\sim$5 m high scarp of delta deposits \citep{Sun2020}. Depending on the wind direction, these topographic features could redirect winds measured by MEDA and Ingenuity.

As mentioned above, the telemetry from flight 59 provides a serendipitous check on the inferred windspeeds. Looking at the data in Figure \ref{fig:Flight_59_Telemetry} between the first two grey regions (between flight times 16 and 22 seconds), the tilt drops to near zero for a short period. During that short window, the $dx/dt$ and $dy/dt$ velocities (pink and purple lines) increase in magnitude. With the yaw holding steady at $-120^\circ$ ($x$ points roughly southwest, and $y$ roughly northwest), the direction of this acceleration corresponds to a drag from a wind blowing in from the north, consistent with the wind azimuth inferred from during the hovers. Assuming a drag coefficient of unity, a drag area equal to the fuselage's area of $0.03\,{\rm m^2}$, an air density of $0.02\,{\rm kg\ m^{-3}}$ \citep{2023JGRE..12807537M}, and a drone mass of $1.8\,{\rm kg}$, the magnitude of this acceleration ($\approx 0.12\,{\rm m\ s^{-2}}$) corresponds to a minimum windspeed of $19\,{\rm m\ s^{-1}}$, which is consistent with the inferred windspeeds. (This value is a ``minimum''  windspeed because Ingenuity's tilt, and therefore wind-ward thrust, does not quite zero out.)

Turning to flight 61 (Figure \ref{fig:Flights_59_and_61_Wind_Profiles}(b)), there appear to be larger variations in windspeed with altitude, but rather than increasing with altitude, as expected for a logarithmic wind profile, the winds increase initially with altitude (on the ascent) before dropping. Then, the one point inferred during descent is smaller even than the predicted windspeed at that altitude. Under unstable conditions on Earth, variations in the horizontal wind tend to fall with altitude \citep{1988aitb.book.....S}, but it is difficult to disentangle altitude from time dependence for the winds for our analysis. The large scatter in tilt azimuth visible in the Flight 61 telemetry (dashed, orange line in Figure \ref{fig:Flight_61_Telemetry}) as compared to Flight 59 suggests more variability in time -- variation in wind azimuth with altitude due, for example, to an Ekman spiral \citep{1992aitd.book.....H} is not expected. Of course, time variability represents an important challenge using a drone to profile the wind for consecutive, rather than simultaneous, measurements, and in Section \ref{sec:Discussion and Conclusions}, we discuss possibilities to address this challenge.

\subsection{Flights 1 through 5}\label{sec:Flights 1 through 5}
\begin{figure}
    \centering
    \includegraphics[width=\textwidth]{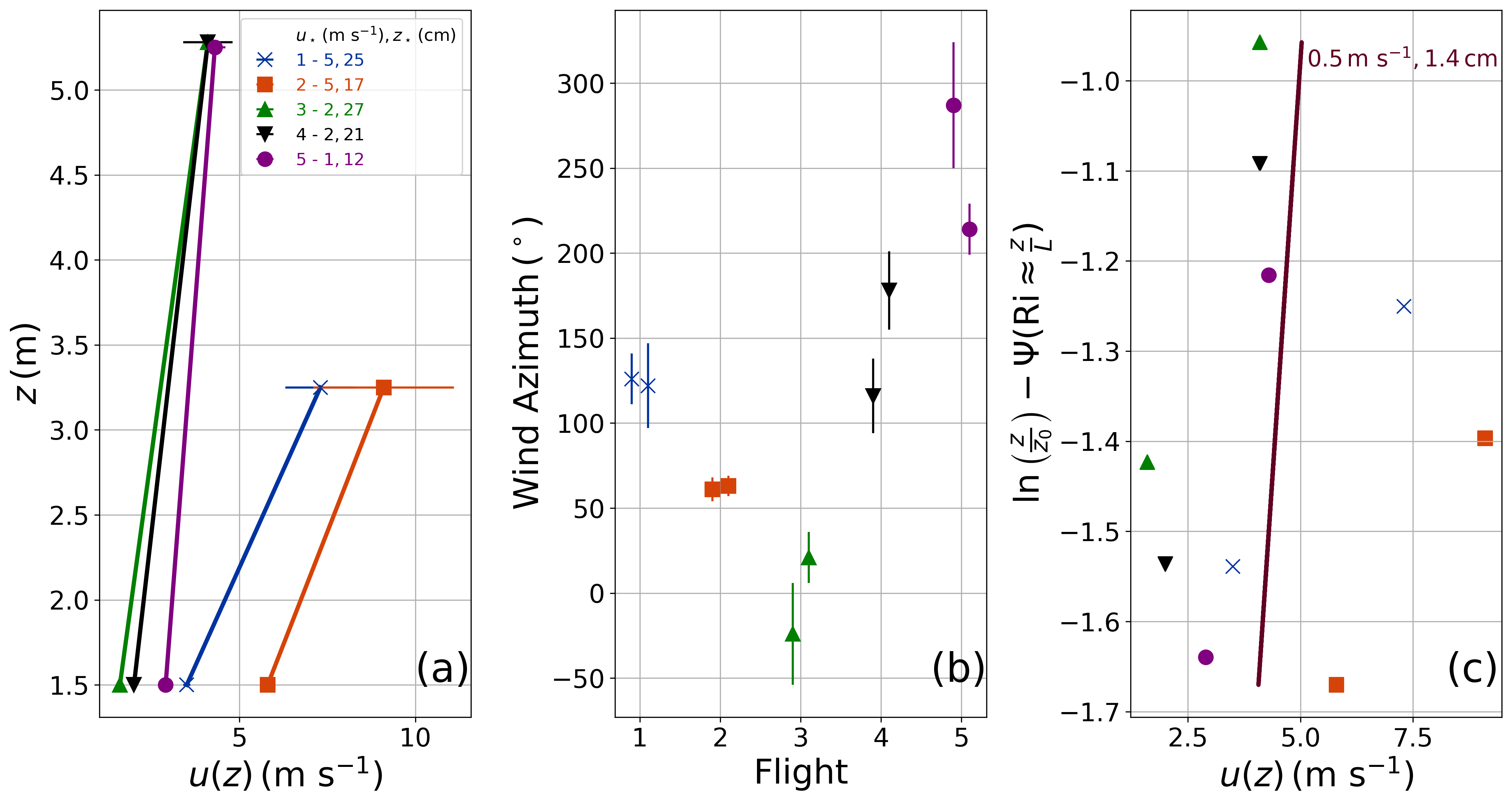}
    \caption{Wind profile analysis using data from during flights 1 through 5. Throughout the three panels, the blue stars show data for flight 1; the orange squares for flight 2; the green, upward-pointing triangles for flight 3; the black, downward-pointing triangles for flight 4; and the purple circles for flight 5. (a) Average windspeeds $u(z)$ measured by MEDA (at $z = 1.5\,{\rm m}$) and inferred from Ingenuity data. The best-fit friction velocity $u_\star$ and roughness length $z_\star$ for each pair of windspeeds are shown in the legend and color-coded by flight. (N.B., we use $z_\star$ instead of the usual $z_0$ for roughness length.) (b) Average wind azimuths and uncertainties from each flight using the same symbols as in panel (a). The azimuth from MEDA is offset just to the left and the azimuth from Ingenuity is offset just to the right of each numbered flight along the x-axis. (c) Wind profile derived by folding all average windspeeds together -- see narrative for an explanation of the $y$-axis. The best-fit $u_\star$ and $z_\star$ are shown.}
    \label{fig:Flights_1-5_Wind_Profiles}
\end{figure}

Figure \ref{fig:Flights_1-5_Wind_Profiles} shows the resulting windspeeds and azimuths from flights 1 through 5, both from MEDA and inferred from Ingenuity telemetry. Starting first with the azimuths, we find good agreement during flights 1 and 2, when the drone was hovering in-place. Azimuths agree less well (to a few tens of degrees) during flights 3 through 5.

Next, we attempted to recover wind profile parameters from average windspeeds inferred from data collected during flights 1 through 5. As discussed in \citet{1988aitb.book.....S}, similarity theory predicts the following near-surface wind profile:
\begin{equation}
    u(z) = \frac{u_\star}{\kappa} \bigg[ \ln \left( \frac{z}{z_\star}\right) - \Psi({\rm Ri}) \bigg],\label{eqn:wind_profile}
\end{equation}
where $u_\star$ is the friction velocity, $\kappa$ is the von K\'{a}rm\'{a}n constant ($\approx 0.4$), $z_\star$ is the roughness scale (N.B., we use $z_\star$ instead of the usual $z_0$ for roughness length), and $\ln$ is the natural log. $\Psi$ is a function of the Richardson number ${\rm Ri}$ and corrects the wind profile for the effects of instability. A combination of analytic and terrestrial field work provides approximate forms for $\Psi$ \citep{2001imm..book.....A}, and we assume the following form widely used in boundary layer studies for both the Earth and Mars \citep{Greeley1995, 2000JGR...10524547S}:
\begin{equation}
    \Psi = 2 \ln \left( \frac{1 + \chi}{2} \right) + \ln\left( \frac{1 + \chi^2}{2} \right) - 2 \tan^{-1} \chi + \frac{\pi}{2}, {\rm Ri < 0}\label{eqn:Psi}
\end{equation}
with $\chi = \left( 1 - 15 {\rm Ri} \right)^{1/4}$. Results from \citet{2023JGRE..12807537M} indicate convective instability during the flights analyzed, implying values of ${\rm Ri} < 0$. It is important to note that Equation \ref{eqn:Psi} is strictly valid in the surface layer, and above about 5-10 m altitude, a mixed layer model should apply \citep{2009JAtS...66.2044M}, which might introduce small corrections. However, with the level of uncertainty involved in this analysis, we do not expect a discernible difference. 

For data collected during flights 1 through 5, we used Equations \ref{eqn:wind_profile} and \ref{eqn:Psi} to explore the wind profiles in several ways, some of which involved linear regression. To that end, we re-cast Equation \ref{eqn:wind_profile} as
\begin{equation}
    y = u(z) = \frac{u_\star}{\kappa} \bigg[ \ln \left( \frac{z}{z_0}\right) - \Psi({\rm Ri}) \bigg] - \frac{u_\star}{\kappa} \ln\left( \frac{z_\star}{z_0} \right) = m x + b,\label{eqn:linear_regression}
\end{equation}
where $z_0$ is NOT the roughness length but is the lowermost height at which winds were measured ($z_0 = 1.5\,{\rm m}$, i.e., MEDA's height). $m$ and $b$ are the slope and intercept, with $x$ defined as the term in square brackets. We can convert $m$ and $b$ into unique values for $u_\star$ and $z_\star$ as

\begin{equation}
    u_\star = \kappa m \label{eqn:u_star_from_slope}, 
\end{equation}
and
\begin{equation}
    z_\star = z_0 e^{-b/m}.\label{eqn:z_star_from_slope_and_intercept}
\end{equation}

Figure \ref{fig:Flights_1-5_Wind_Profiles}(a) illustrates our first approach to exploring wind profiles. For the data collected during each flight, there are data points at only two heights -- one from MEDA and one from Ingenuity -- which suffices to determine a line. The figure illustrates the best-fit wind profile and $u_\star$ and $z_\star$ for each individual data set. With two points, uncertainties on each are quite large, $\sigma_{u_\star} \sim 1\,{\rm m\ s^{-1}}$ and $\sigma_{z_\star} \sim z_\star$. By comparison, for the times and locations of flights 1 through 5, \citet{2023JGRE..12807537M} estimated $u_\star$ between $0.368$ and $0.430\,{\rm m\ s^{-1}}$ and $z_\star = 1\,{\rm cm}$, much smaller than our estimates. Indeed, the lack of observed dust and sand motion in the Mars 2020 video of flight 1 suggests $u_\star$ did not exceed the threshold for aeolian transport, $\lesssim 1\,{\rm m\ s^{-1}}$ \citep{NEWMAN2022637}.

One obvious way to improve constraints on the wind profile parameters is to add more sample points. Given that the $u_\star$ values estimated by \citet{2023JGRE..12807537M} range by less than 15\% (and $z_\star$ is assumed constant), for our second approach to exploring wind profiles, we combined windspeed estimates from flights 1 through 5 into one wind profile, as shown in Figure \ref{fig:Flights_1-5_Wind_Profiles}(c). Determining the $y$-value for points in the plot requires an estimate of ${\rm Ri}$ for each point. For those estimates, we used the Monin-Obukhov length $L$ estimated by \citet{2023JGRE..12807537M} for the times and locations of flights 1 through 5 -- these values ranged from $-0.796$ to $-0.524\,{\rm m}$, the negative values indicating convective instability. Then, we approximated ${\rm Ri} \approx z/L$ (cf. p.~220 of \citealp{2001imm..book.....A}), with $z$ the height of either MEDA (1.5 m) or Ingenuity (3.25 or 5.28 m, depending on the flight). With that approximation, we stacked all the datapoints from during all the flights together. The best-fit $u_\star$ and $z_\star$ (shown in the figure) more closely match the values from \citet{2023JGRE..12807537M}, although the lack of an obvious linear trend means the formal uncertainties on these best-fit estimates are, again, very large. (Since we have no uncertainties for $L$, we do not consider uncertainties for this wind profile.) We did not attempt to combine profile points for flights 59 and 61 since they took place at locations separated by hundreds of meters and at different times of day, so we would not necessarily expect they reflect similar aerodynamic conditions.

\begin{figure}
    \centering
    \includegraphics[width=\textwidth]{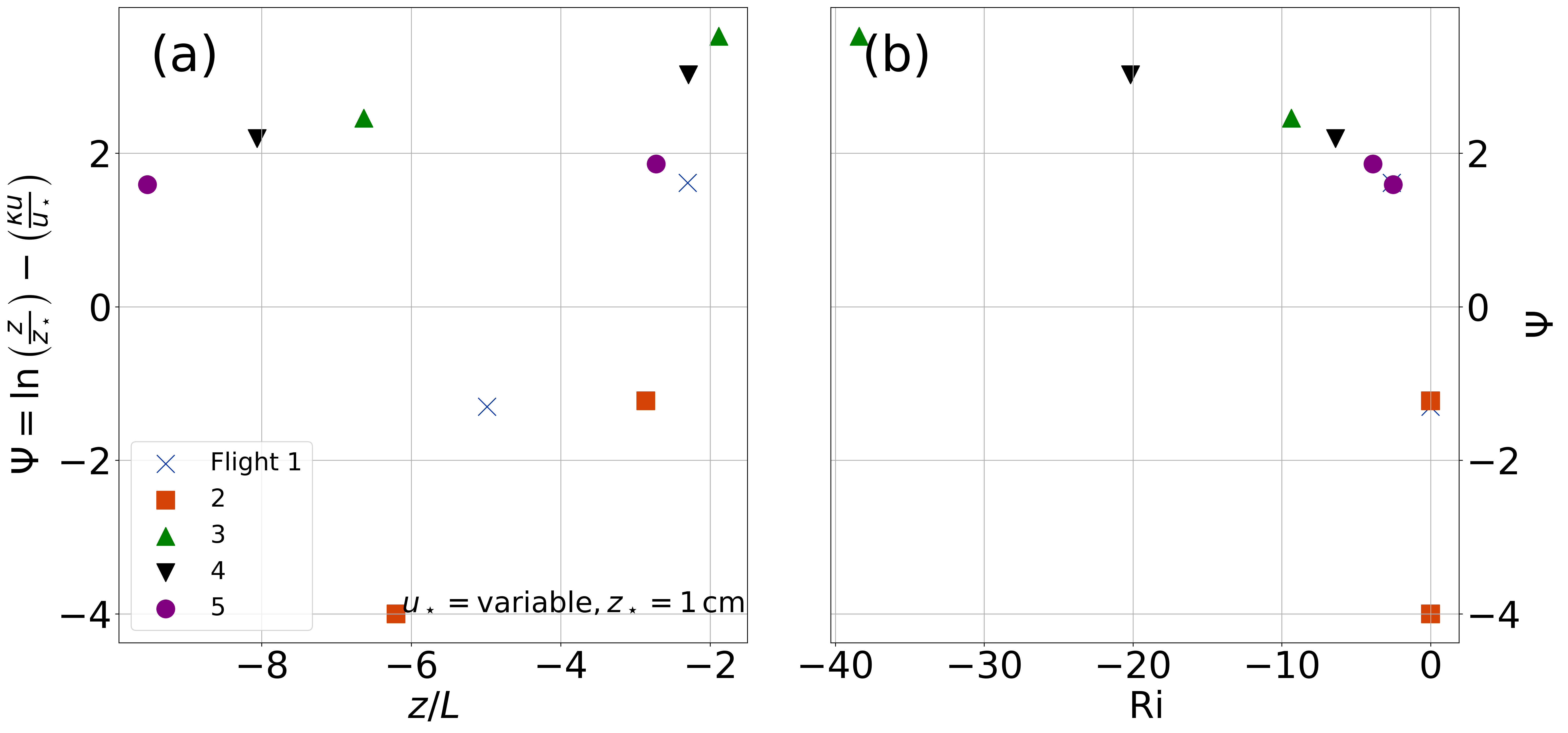}
    \caption{Exploring the best-fit Richardson numbers ${\rm Ri}$. (a) For each data pair from during all flights, the inferred $\Psi$-value vs.~the approximate ${\rm Ri} \approx z/L$, where $z$ is the height at which the average windspeed is estimated and $L$ is the inferred Monin-Obukhov length. For this plot, we assumed a $u_\star$ for each pair of data points derived from MEDA data and $z_\star = 1\,{\rm cm}$ \citep{2023JGRE..12807537M}. (b) The best-fit ${\rm Ri}$ ($x$-axis) as derived from the $\Psi$-values shown on the y-axis of panel (a). Negative $\Psi$-values are unphysical and are plotted with ${\rm Ri} = 0$. In both panels, the symbols correspond to the same flights as in Figure \ref{fig:Flights_1-5_Wind_Profiles}.}
    \label{fig:Flights_1-5_Psi_vs_Ri}
\end{figure}

We can also flip the analysis around and explore what ${\rm Ri}$ values are implied by the data, and this analysis is illustrated in Figure \ref{fig:Flights_1-5_Psi_vs_Ri}. For this analysis, we assumed the $u_\star$ and $z_\star$ values appropriate for each flight from \citet{2023JGRE..12807537M}. Panel (a) shows $\Psi$ implied by solving Equation \ref{eqn:wind_profile} (plotted against the approximate ${\rm Ri}$). Most of the data points correspond to physically reasonable values, while three points (one from Ingenuity during flight 1 and both data points from during flight 2) have unphysical negative values. This result suggests that the correct $u_\star$ is larger than assumed (which would reduce $\kappa u/u_\star$ and allow $\Psi$ to be positive), the appropriate $z_\star$ is much smaller than $1\,{\rm cm}$, or that the inferred $u$-values are too large. This latter possibility seems unlikely since, for flight 2 (orange squares), the windspeed from both Ingenuity and from MEDA would have to be too large. 

\section{Discussion and Conclusions}
\label{sec:Discussion and Conclusions}

How should we interpret the large windspeeds implied by Ingenuity's telemetry? Without concurrent MEDA windspeed data against which to compare for flights 59 and 61, the most we can easily say is that models somewhat under-predicted the local windspeeds and predicted wind azimuths different from those observed. These results comport somewhat with inference of higher-altitude (2 km) windspeeds during Perseverance's parachute descent -- wind directions were correctly predicted, but windspeeds were under-predicted \citep{2024Icar..41516045P}. Since they have concurrent MEDA wind data, flights 1 through 5 present a more interesting challenge.


Given the relatively short hover times at altitude, one obvious possibility is that the winds seen by Ingenuity simply represent a natural excursion around a smaller average value. We discuss the challenge of diagnosing wind profiles by drone in the presence of variability below, but a simple statistical approach argues that the Ingenuity-inferred winds are probably not the result of natural variability. 

We can use the first term in Equation \ref{eqn:wind_profile} to scale average MEDA-observed windspeeds up to the expected average values at Ingenuity's altitudes. (Including $\Psi$ reduces the expected windspeeds and makes the large windspeeds inferred from Ingenuity telemetry even more unlikely, as discussed below.) In addition to $u_\star$, $z_\star$, and Monin-Obuhkov lengths, \citet{2023JGRE..12807537M} provided estimates for windspeeds from MEDA averaged over half-hour blocks, and for this analysis, we take those averages centered on the times for each of Ingenuity's flights. For example, during flight 1, this approach predicts an average windspeed at 3.25 m altitude of $6.0\,{\rm m\ s^{-1}}$. The natural variability in windspeeds is often described using a Weibull distribution \citep[][and references therein]{1996JSpRo..33..754L, 2022JGRE..12707523V}, which involves a shape parameter $k$ and a scale parameter $c$ related to the median of the distribution $m$ as $m = c \times \left( \ln 2 \right)^k$. Based on MEDA data, \citet{2022JGRE..12707523V} estimated $k=1.49$ at Jezero. Using the scaled windspeeds to estimate the appropriate scale parameter for each flight, we can estimate the probability, assuming a Weibull distribution, for Ingenuity to have seen the windspeed it did see, plus-and-minus the reported uncertainties. So, taking flight 1 as our example again, we have a probability of about 15\% for the windspeed at 3.25 m to have lain between 6.3 and $8.3\,{\rm m\ s^{-1}}$. Similarly small probabilities obtained for all the flights, with the largest probability for flight 2 (22\%). Conservatively assuming this maximum probability applied to all five flights, we would only have expected to see one flight with the large windspeeds Ingenuity did see. 

\begin{figure}
    \centering
    \includegraphics[width=\textwidth]{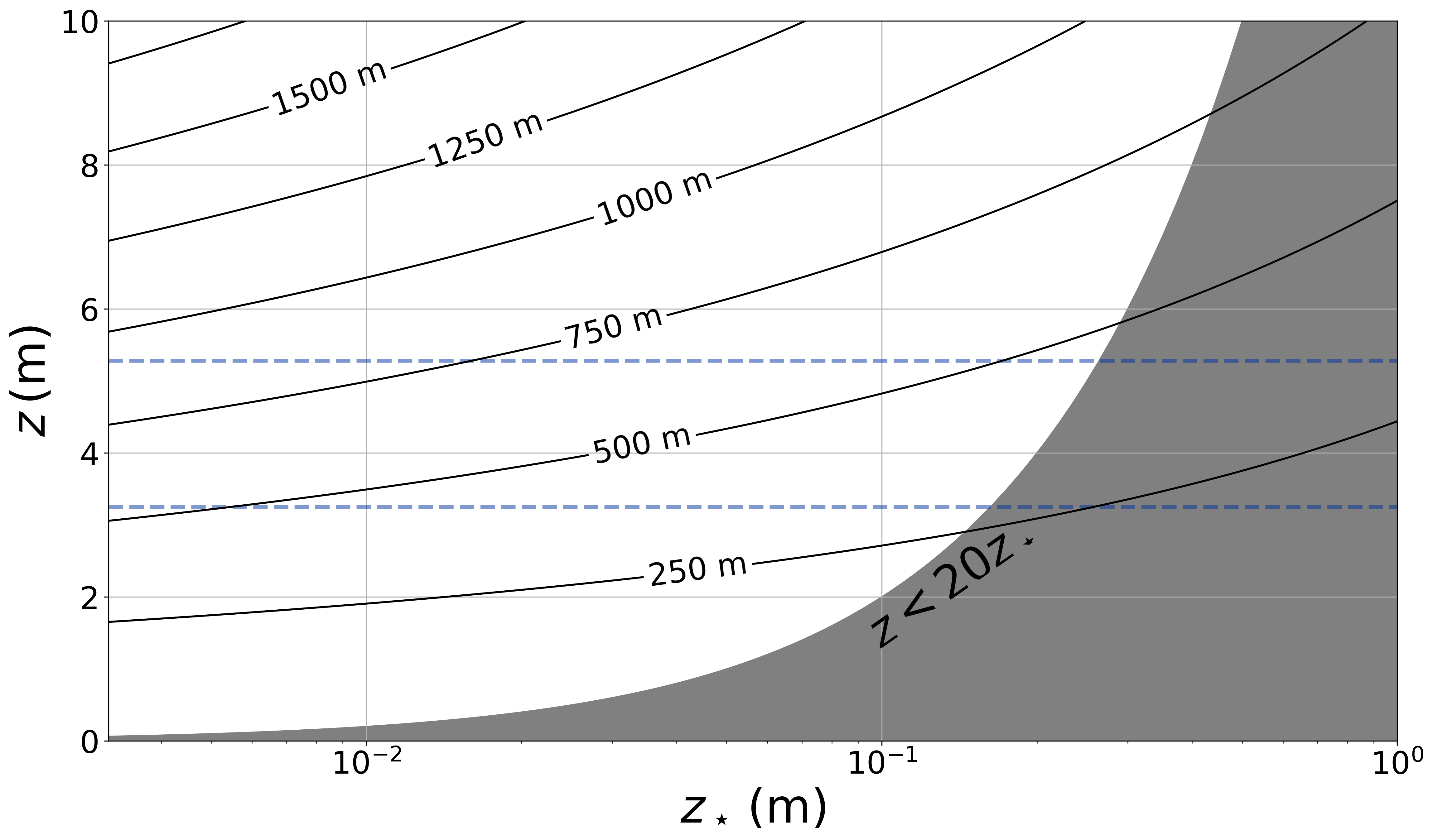}
    \caption{Contours of upwind fetch distance to which the wind profile measured at height $z$ over a surface of roughness length $z_\star$ is sensitive. The dashed blue lines show $z = 3.25$ and $5.28\,{\rm m}$, the altitudes at which Ingenuity probed winds during flights 1-5. The shaded grey regions shows where $z < 20 z_\star$, heights low enough that the logarithmic wind profile may not apply and the flow becomes laminar.}
    \label{fig:Fetch_Distance}
\end{figure}

Instead, the large windspeeds may reflect the aerodynamic environment far upwind from Ingenuity. In the course of a comprehensive discussion about constraining wind profiles from field measurements, \citet{1993BoLMe..63..323W} pointed out the wind profile probed at a given height above a point on the surface will reflect the aerodynamic conditions upwind from that point. The higher the height probed, the farther upwind. This upwind fetch distance therefore depends both on the roughness length $z_\star$ and the height at which the profile is probed $z$. Based on Equation 5 in \citet{1993BoLMe..63..323W}, Figure \ref{fig:Fetch_Distance} shows the upwind fetch distance as a function of $z_\star$ and $z$. 
At the height of MEDA's wind sensor (1.5 m), the fetch distance is less than 200 m. By contrast, for Ingenuity's altitudes, the upwind fetch distance is nearly 500 m (assuming $z_\star \geq 1\,{\rm cm}$). Taking flight 1 as our example again, winds came in from the southeast. The terrain immediately surrounding Mars 2020 and Ingenuity is relatively smooth in appearance, as seen in the available HiRISE imagery. Looking 500 m to the southeast, the terrain appears much rougher, with sharp ridges and craters prominent \citep{2020SSRv..216..127S}. Indeed, the terrain appears to roughen in all directions outward from Ingenuity's location during flights 1-5. Thus, it seems likely that the large windspeeds seen during these flights reflect a significantly different aerodynamic environment from the environment very near Mars 2020. This result highlights a key consideration for determining wind profile parameters -- using measurements taken from a few to many meters height, the parameters may not reflect the immediately local conditions. 

The winds measured by Ingenuity appear to be consistent with the pattern of inferred winds at the time of Perseverance's landing ($L_{\rm s} = 5^\circ$, 15:15 LTST) obtained by \citet{2024Icar..41516045P}. The inferred windspeed and direction were obtained by analyzing the displacement of the jettisoned backshell-parachute on the surface. Easterly winds, with an average speed of $11\,{\rm m\ s^{-1}}$ over altitudes of 2 km to the surface, were found to have blown the backshell-parachute towards the west. Closer to the surface, MEDA observations, obtained after $L_{\rm s} = 22^\circ$, indicated a wind blowing from the east at a speed of $7\,{\rm m\ s^{-1}}$. The apparent increase in wind speed with altitude would then be consistent with Ingenuity's flight 61 (sol 933) and expected from theory when extrapolating the wind profile upwards.

\citet{2024Icar..41516045P} also inferred the windspeed and direction close to the surface by tracking a dust cloud that was kicked up by Perserverance’s heat shield when it impacted the surface. The dust cloud's direction was towards the north, presumably blown by southerly winds, and had a speed of $5\,{\rm m\ s^{-1}}$. In contrast, MEDA observations, on subsequent sols, indicated that the prevailing wind direction was from the east. In terms of magnitude, this difference in wind direction is consistent with differences between wind directions measured by Ingenuity and MEDA during flights 3, 4, and 5 (sols 64, 69, 76).

In any case, our results represent the first attempt to probe the martian boundary layer at near-surface altitudes using a drone. As such, they establish a baseline for future efforts. Unfortunately, Ingenuity suffered catastrophic damage to its rotor blades in 2024 Jan, foreclosing additional attempts to profile boundary layer winds. However, Ingenuity did complete 72 flights total, and the telemetry from all the flights may soon be made available for additional analysis. An improved understanding of the aerodynamic response of the helicopter will be critical to wringing high quality science out of these data.

Although this study is the first wind profile on Mars to use a drone, it is not the first wind profile attempt. The Imager for Mars Pathfinder (IMP) windsock experiment involved hanging three pendula at 0.53, 0.82, and 1.1 m height above the ground. The pendula would settle at an angle related to the windspeed. \citet{2000JGR...10524547S} analyzed images of the windsocks and estimated the windspeeds at each height for several different sols. Unfortunately, aerodynamic interference from the lander body muddled data from the lowermost windsock in most analyses, resulting in two sample points during all but one reported analysis. \citet{2000JGR...10524547S} reported $u_\star \approx 1\,{\rm m\ s^{-1}}$ and $z_\star \approx 3\,{\rm cm}$ for all their wind profiles, consistent with expectations. However, the formal uncertainties on these estimates were actually quite large. Inspection of their Figure 7 shows uncertainties on the inferred windspeeds, $\sigma \sim 1\,{\rm m\ s^{-1}}$. By re-analyzing the sol 52 data (the only wind profile with three points) and folding in these uncertainties, we found $u_\star = \left( 0.94\pm0.54 \right)\,{\rm m\ s^{-1}}$ and $z_\star = \left( 3\pm5 \right)\,{\rm cm}$, meaning $u_\star$ was estimated to limited precision and $z_\star$ could only be inferred as less than $8\,{\rm cm}$. (Following on the previous discussion, a measurement at 1.1 m height with $z_\star = 3\,{\rm cm}$ has a fetch distance of about 100 m.)

Such large uncertainties are not surprising, given the large uncertainties on the windspeeds and the small number of sample points $N$. We can explore how the uncertainties on profile parameters depend on these details to frame how best to make such measurements. Consider a wind profile given only by the first term in the square brackets in Equation \ref{eqn:wind_profile} and assume a number of profile points $N$ spaced logarithmically \citep{1993BoLMe..63..323W} such that the altitudes of adjacent points $\delta x \equiv \ln \left( z_{i + 1}/z_i \right) = {\rm const.}$, and a typical uncertainty on the average windspeed $\sigma$. The expected uncertainty on $u_\star$ and $z_\star$ can be estimated analytically as \citep{Diniega2024}:
\begin{equation}
\left( \frac{\sigma_{u_\star}}{\sigma} \right) \approx \kappa \sqrt{ \frac{12}{N \left( N^2 - 1 \right) \delta x^2 } },\label{eqn:fractional_ustar_uncertainty}
\end{equation}
and 
\begin{equation}
\frac{\sigma_{z_\star}}{z_\star} \approx \kappa \left( \frac{\sigma}{u_\star} \right) \ln\left( \frac{z_0}{z_\star} \right) \sqrt{ \frac{12}{N \left( N^2 - 1 \right) \delta x^2 } }.\label{eqn:fractional_zstar_uncertainty}
\end{equation}
It is worth noting that these equations do not depend on the manner in which the wind profile is estimated. They apply even in the case of simultaneous measurements by a series of anemometers mounted along a mast (as is often done on the Earth -- \citealp{2023PSJ.....4..102Z}).

\begin{figure}
    \centering
    \includegraphics[width=\textwidth]{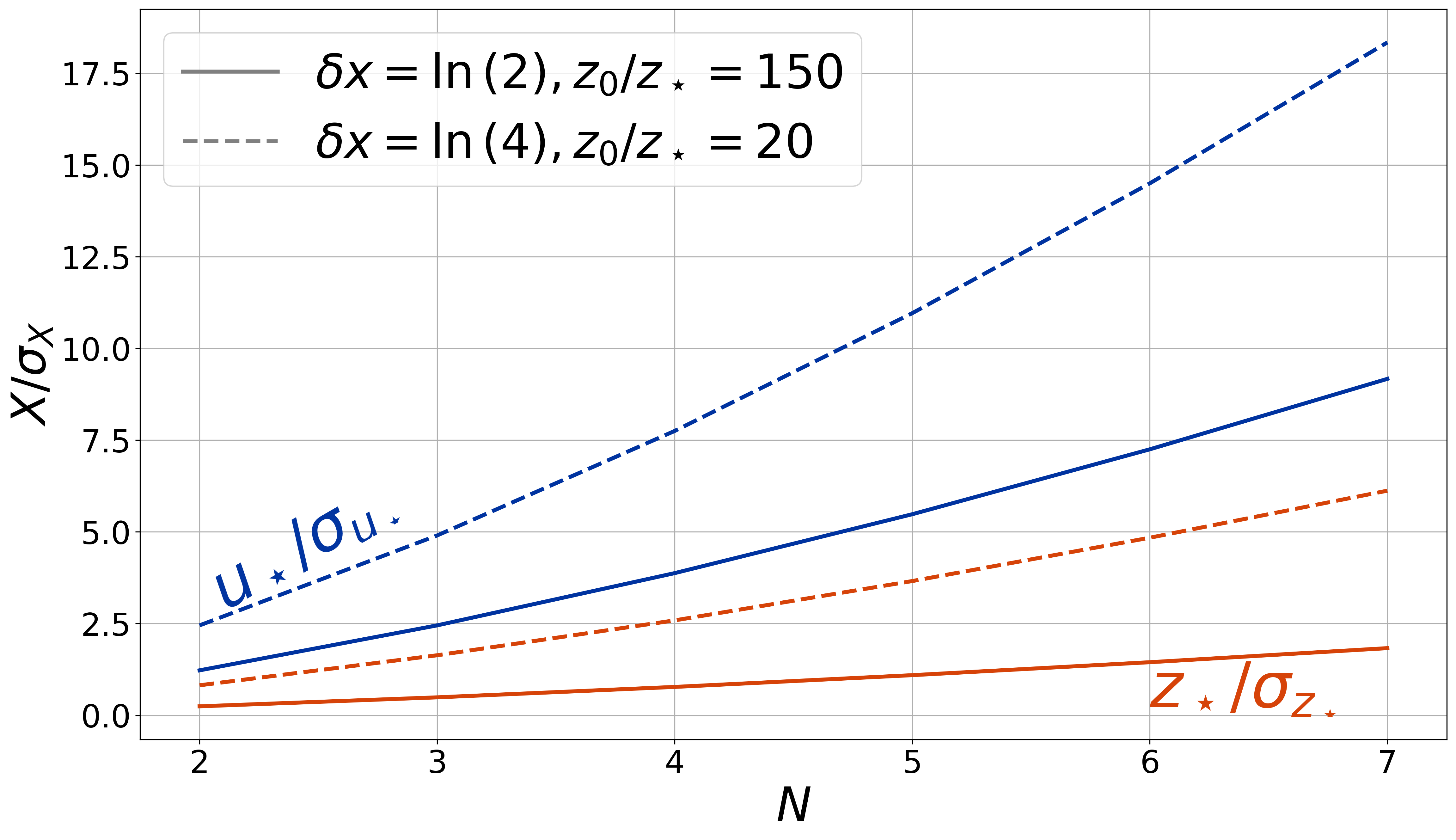}
    \caption{Dependence of wind profile parameter uncertainties for $u_\star$ and $z_\star$, given by $\sigma_{u_\star}$ and $\sigma_{z_\star}$, respectively. $N$ is the number of points sampled along the wind profile; $\delta x \equiv \ln \left( z_{i+1}/z_i\right)$, i.e., the spacing of sample points; and $z_0/z_\star$ is the ratio of the height of the lowermost sample point to the roughness length. Blue lines (solid and dashed) $u_\star/\sigma_{u_\star}$, while orange lines show $z_\star/\sigma_{z_\star}$. We have assumed $\sigma/u_\star = 1$ for these calculations. The solid lines illustrate the conditions closest to those analyzed in this study.}
    \label{fig:X_over_sigma_X}
\end{figure}

These equations illustrate several key dependencies. Not surprisingly, the more sample points collected along the profile $N$, the smaller the uncertainties on both parameter estimates. The spacing between profile points reflected in $\delta x$ also impacts the uncertainties -- the more widely spaced the points (i.e., larger $\delta x$), the smaller the uncertainties because, in principle, the expected difference in windspeeds from one point to the next is larger, helping better constrain the profile. Figure \ref{fig:X_over_sigma_X} shows these results. (To simulate the analyses presented here, we have assumed $\sigma/u_\star = 1$ for these calculations.)

Figure \ref{fig:X_over_sigma_X} also illustrates another key result. Whether $z_\star$ can be accurately determined for a given $N$ depends sensitively on the height of the lowermost profile point, $z_0$. For our analysis of the Ingenuity and MEDA data, $z_0/z_\star \approx 1.5\,{\rm m}/1\,{\rm cm} = 150$, a situation represented by the solid lines in Figure \ref{fig:X_over_sigma_X}. In that case, even seven points along the profile would not have resulted in a $z_\star$ estimate with high confidence. \citet{1993BoLMe..63..323W} suggested not probing below a few tens of $z_\star$ since, too close to the surface, the wind profile becomes laminar. Figure \ref{fig:X_over_sigma_X} shows that, for $z_0/z_\star = 20$ (i.e., $z_0 = 20\,{\rm cm}$ for $z_\star = 1\,{\rm cm}$), five sample points along the profile would have provided $z_\star/\sigma_{z_\star} \approx 3$. Hovering Ingenuity at such a low altitude would likely have been hazardous. These results illustrate the difficulty of recovering $z_\star$, even in the ideal case of a stationary wind field. 

But, as already pointed out, the winds can be quite variable near a planetary surface. Indeed, the power-spectrum of near-surface winds can be an important diagnostic of energy exchange and dissipation in the boundary layer \citep{2011RvGeo..49.3005P}. Although theoretical expectations suggest a power-spectrum that declines toward higher frequencies with an index of -5/3 \citep{1991RSPSA.434...15K}, previous analyses of near-surface winds on Mars suggested a slightly shallower slope of -1.36 \citep{2022JGRE..12707523V}. The high sampling rate for Ingenuity's telemetry (500 Hz) means it may be an excellent platform for high-frequency sampling of the martian boundary layer, and, in fact, terrestrial field experiments show good agreement in the power-spectra between directly measured and drone-inferred winds \citep{2024JAtOT..41...25B}. Here, again, a robust understanding of Ingenuity's aerodynamic response would be prove very helpful.

In this connection, we note that \citet{grip2020modeling} has simulated the response of Ingenuity to a $5\, {\rm m\ s^{-1}}$ step gust. Although the MEDA data show no evidence for such rapid changes in winds during the flight phases we analyzed, the results from \citet{grip2020modeling} show that the inflow dynamics of the rotor during a large gust produce a pitch moment away from the gust such that Ingenuity would be expected to exhibit a transient response more complex than the simple steady-state, trimmed flight condition we've assumed here. Assessing the impact of this effect on windspeed estimates requires a fuller investigation of the gust response of the vehicle explicitly, but this effect may give rise to overestimated windspeeds.

The variability represents a significant challenge to probing a wind profile by drone for two additional reasons: (1) winds may change over time as the drone samples the different altitudes consecutively, as compared to an anemometer tree via which winds are measured at all points simultaneously; and (2) energy constraints during hover limits the duration over which winds can be averaged to tens of seconds, as compared to the tens of minutes over which wind profiles are typically averaged. Hovering longer at a particular altitude can reduce the uncertainty on the average windspeed inferred for that altitude but at the cost of increasing the time offset for averages for multiple altitudes. 

As a preliminary exploration of that issue, we generated tens of thousands of synthetic datasets meant to replicate the 500-Hz point-to-point windspeeds inferred from Ingenuity's telemetry. We assumed $u_\star = 0.3\,{\rm m\ s^{-1}}$ and $z_\star = 1\,{\rm cm}$ and calculated synthetic wind profiles sampled at two points and at five points in consecutive time windows. We included point-to-point scatter and red noise meant to replicate the structure of the noise in the actual windspeed data. We found that the wind profile parameter estimates from averaging over 20-second windows had nearly the same precisions as estimates involving averages over 30-minute windows, with the five-sample point profiles consistently achieving better recovery of the parameters. Future work should more thoroughly explore these issues, but these preliminary results show the promise of drones for profiling winds.

Of course, drones may furnish opportunities for more than just wind profiling on Mars. Measuring water vapor with altitude throughout the day (and perhaps during the night) could provide insights into the exchange of water between the surface and sub-surface, key for understanding the martian water cycle \citep{2017acm..book..295M}. The threshold windspeeds for dust and sand mobilization remain highly uncertain for Mars, severely hampering our understanding of the martian dust cycle \citep{NEWMAN2022637}. Here, preliminary work using images of dust lifted by Ingenuity's prop wash has provided useful constraints \citep{2022JGRE..12707605L}, but limits on our knowledge of the shear stresses generated by the prop wash gave rise to significant uncertainties. Measuring the temperature profile in the martian surface layer would also clarify the implications of the greater contribution of the radiative heat flux relative to the sensible heat flux on Mars for applying similarity theory to Mars \citep{2017acm..book..106R}. This work would improve model parameterizations of convection and therefore the vertical wind and temperature profiles in the lower atmosphere. Better calibration of Ingenuity's (or a future drone's) wind stresses could yield foundational aeolian experiments for Mars. And such boundary layer experiments could be conducted as part of a drone-based mission with a broader focus. For instance, \citet{2023PSJ.....4..155M} explored options for mapping magnetic signatures in geological strata across Mars via a drone-based mission, and \citet{2024LPICo3007.3350F} discussed prospects for a mission to Vallis Marineris. Such a mission would, in the course of traveling from waypoint to waypoint, inadvertently collect a wealth of wind data through its telemetry. The same holds true for any drone-based mission on any world. Thus, as prospects for exploring the solar system by drone grow brighter, exploration of near-surface planetary boundary layers will become not only more scientifically fruitful but also inevitable.

\begin{acknowledgments}
The authors acknowledge thoughtful input from two anonymous referees. BJ acknowledges support from NASA's Solar System Workings program, grant 17-SSW17-0094, support from NASA's Mars Data Analysis Program, grant 22-MDAP22\_2-0012, and help from the NASA PDS Atmospheres Node, especially Dr.~Lynn Neakrase. A portion of this work was done at the Jet Propulsion Laboratory, California Institute of Technology under a contract with the National Aeronautics and Space Administration (80NM0018D0004). AM is supported by the grant PRE2020-092562 funded by MCIN/AEI/10.13039/501100011033 and by ``ESF Investing in your future''. CN is supported by MEDA funding. RL is supported by the Mars 2020 project SuperCAM investigation via a JPL contract. The authors also acknowledge useful feedback from Ricardo Hueso and Agustin Sanchez Lavega. 
\end{acknowledgments}

\software{astropy \citep{astropy:2013, astropy:2018, astropy:2022}, colorednoise, emcee \citep{2013PASP..125..306F}, george \citep{2015ITPAM..38..252A}, iPython \citep{PER-GRA:2007}, numpy \citep{harris2020array}, matplotlib \citep{Hunter:2007}, scipy \citep{2020SciPy-NMeth}}

\bibliography{sample631}{}
\bibliographystyle{aasjournal}



\end{document}

%% file: Windspeeds_table.tex



\begin{deluxetable}{ccccccccccc}


\tabletypesize{\scriptsize}


\tablecaption{Wind Data}

\tablenum{1}

\tablehead{\colhead{Flight} & \colhead{Sol} & \colhead{$L_{\rm s}$} & \colhead{Time of Day} & \colhead{Long.~N} & \colhead{Lat.~E} & \colhead{MEDA Windspeed} & \colhead{MEDA Az} & \colhead{Ingenuity Alt} & \colhead{Ingenuity Windspeed} & \colhead{Ingenuity Az} \\ 
\colhead{} & \colhead{} & \colhead{($^\circ$)} & \colhead{LTST/LMST} & \colhead{($^\circ$)} & \colhead{($^\circ$)} & \colhead{(${\rm m\ s^{-1}}$)} & \colhead{($^\circ$)} & \colhead{(${\rm m}$)} & \colhead{(${\rm m\ s^{-1}}$)} & \colhead{($^\circ$)} } 

\startdata
1 &   58 &   33.67268 &  12:13/12:30 &  18.44486 &  77.45102 &  $3.5\pm0.1$ &  $126\pm15$ &  3.25 &          $7.3\pm1.0$ &   $122\pm25$ \\
2 &   61 &   35.07378 &  12:14/12:30 &  18.44486 &  77.45102 &  $5.8\pm0.1$ &  $61\pm7$ &   5.28 &          $9.1\pm2.0$ &   $63\pm6$ \\
3 &   64 &   36.47081 &  12:15/12:30 &  18.44486 &  77.45101 &  $1.6\pm0.1$ &  $336\pm30$  &  5.28 &          $4.1\pm0.7$ &   $21\pm15$ \\
4 &   69 &   38.79059 &  12:17/12:30 &  18.44486 &  77.45112 &  $2.0\pm0.1$ &  $116\pm22$&  5.28 &          $4.1\pm0.7$ &   $178\pm23$ \\
5 &   76 &   42.02140 &  12:19/12:30 &  18.44267 &  77.45139 &  $2.9\pm0.1$ &  $287\pm37$ &  5.28 &          $4.3\pm0.3$ &   $214\pm15$ \\
59 &  915 &  108.2954 &  11:25/11:02 &  18.48940 &  77.34834 &  - &            - &       4.25  (up) &    $18.6\pm3.8$ &  $2\pm5$ \\
\ditto &  \ditto &  \ditto &    \ditto &                 \ditto &    \ditto &    - &            - &       8.25  (up) &    $19.8\pm2.3$ &  $3\pm5$ \\
\ditto &  \ditto &  \ditto &    \ditto &                 \ditto &    \ditto &    - &            - &       12.25 (up) &    $22.9\pm1.7$ &  $0\pm10$ \\
\ditto &  \ditto &  \ditto &    \ditto &                 \ditto &    \ditto &    - &            - &       16.25 (up) &    $24.3\pm1.2$ &  $20\pm6$ \\
\ditto &  \ditto &  \ditto &    \ditto &                 \ditto &    \ditto &    - &            - &       20.25 (up) &    $22.6\pm1.4$ &  $11\pm7$ \\
\ditto &  \ditto &  \ditto &    \ditto &                 \ditto &    \ditto &    - &            - &       16.25 (down) &  $18.4\pm0.8$ &  $8\pm8$ \\
\ditto &  \ditto &  \ditto &    \ditto &                 \ditto &    \ditto &    - &            - &       12.25 (down) &  $22.0\pm1.5$ &  $-3\pm4$ \\
\ditto &  \ditto &  \ditto &    \ditto &                 \ditto &    \ditto &    - &            - &       8.25  (down) &  $22.7\pm0.4$ &  $-11\pm4$ \\
\ditto &  \ditto &  \ditto &    \ditto &                 \ditto &    \ditto &    - &            - &       4.25  (down) &  $23.3\pm0.5$ &  $-13\pm5$ \\
61 &  933 &  116.5606 &  12:28/12:02 &  18.49392 &  77.34511 &  - &            - &       5.28  (up) &    $7.8\pm0.9$ &   $11\pm9$ \\
\ditto &  \ditto &  \ditto &    \ditto &                 \ditto &    \ditto &    - &            - &       10.28 (up) &    $10.1\pm1.0$ &  $15\pm10$ \\
\ditto &  \ditto &  \ditto &    \ditto &                 \ditto &    \ditto &    - &            - &       15.28 (up) &    $13.8\pm1.6$ &  $-4\pm10$ \\
\ditto &  \ditto &  \ditto &    \ditto &                 \ditto &    \ditto &    - &            - &       20.28 (up) &    $12.5\pm1.2$ &  $-7\pm11$ \\
\ditto &  \ditto &  \ditto &    \ditto &                 \ditto &    \ditto &    - &            - &       24.28 (up) &    $7.0\pm0.6$ &   $-19\pm13$ \\
\ditto &  \ditto &  \ditto &    \ditto &                 \ditto &    \ditto &    - &            - &       10.28 (down) &  $4.4\pm0.7$ &   $-29\pm13$ \\
\enddata


\tablecomments{``Long.~N/Lat.~E'' are the location of Ingenuity's take-off point, longitude measured north and latitude measured east. ``MEDA/Ingenuity Windspeed'' are the MEDA- or Ingenuity-inferred windspeed, averaged over the chosen flight window. ``MEDA/Ingenuity Az'' are the MEDA- or Ingenuity-inferred incoming wind azimuth. The ``Ingenuity Altitude'' column also notes whether the windspeed estimate comes from the ascent (``up'') or descent (``down'').}
\label{tbl:Wind_Data}


\end{deluxetable}